\documentclass[aps,pra,showpacs,floatfix,twocolumn,superscriptaddress]{revtex4-2}
\usepackage[utf8]{inputenc}
\usepackage[]{graphicx}
\usepackage{subcaption}
\usepackage[singlelinecheck=false, justification=justified]{caption}
\usepackage{epstopdf}
\usepackage{dcolumn}% Align table columns on decimal point
\usepackage{bm}% bold math
\usepackage{siunitx}
\usepackage{enumerate}
\usepackage{lineno}
\usepackage{physics,amssymb,amsmath}

\newcommand{\hp}{\hat{p}}
\newcommand{\hx}{\hat{x}}
\newcommand{\hS}{\hat{S}_x}
\newcommand{\ha}{\hat{a}}
\newcommand{\had}{\hat{a}^\dagger}

\newcommand{\rev}[1]{{\color{black}{#1}}}

\usepackage[dvipsnames]{xcolor}

\begin{document}

%\title{Table-top nanodiamond interferometer towards revealing quantum features in gravity: a feasibility study}
\title{Table-top nanodiamond interferometer enabling quantum gravity tests}
\author{Marta Vicentini}
\thanks{These authors have equally contributed to this work.}
\affiliation{INRIM, Strada delle Cacce 91, I-10135 Torino, Italy}
\author{Matteo Bordin}
\thanks{These authors have equally contributed to this work.}
%\affiliation{INRIM, Strada delle Cacce 91, I-10135 Torino, Italy}
\affiliation{Queen's University, Belfast BT7 1NN, United Kingdom}
\author{Ettore Bernardi}
%\email{e.bernardi@inrim.it}
\affiliation{INRIM, Strada delle Cacce 91, I-10135 Torino, Italy}

\author{Ekaterina Moreva}
\affiliation{INRIM, Strada delle Cacce 91, I-10135 Torino, Italy}
\author{Fabrizio Piacentini}
\email{f.piacentini@inrim.it}
\affiliation{INRIM, Strada delle Cacce 91, I-10135 Torino, Italy}
\author{Carmine Napoli}
\affiliation{INRIM, Strada delle Cacce 91, I-10135 Torino, Italy}
\author{Ivo Pietro Degiovanni}
\affiliation{INRIM, Strada delle Cacce 91, I-10135 Torino, Italy}
\affiliation{INFN, sezione di Torino, via P. Giuria 1, 10125 Torino, Italy}
\author{Alessandra Manzin}
\affiliation{INRIM, Strada delle Cacce 91, I-10135 Torino, Italy}
\author{Marco Genovese}
\affiliation{INRIM, Strada delle Cacce 91, I-10135 Torino, Italy}
\affiliation{INFN, sezione di Torino, via P. Giuria 1, 10125 Torino, Italy}
\date{\today}

\begin{abstract}

%Unifying quantum theory and general relativity is the dream of contemporary physics.
%Recently, it has been argued that the observation of gravity-induced entanglement could demonstrate the quantum nature of gravity; a few experimental proposals have been advanced, but extreme technological requirements make their implementation still far.
%We present a feasibility study for a table-top interferometer enabling less demanding quantum gravity tests.
%By relying on quantum superpositions of steady massive objects, our interferometer requires short-range time-varying magnetic fields (much easier to implement) and allows re-using the quantum probes (inevitably lost in previous schemes).

Along the path of unifying quantum theory and general relativity, it has recently been argued that the observation of gravity-induced entanglement could demonstrate the quantum nature of gravity; a few experimental proposals have been advanced, but extreme technological requirements make their implementation still far.
Here we propose a table-top interferometer that, by relying on quantum superpositions of steady nanodiamonds, enables less demanding quantum gravity tests requiring short-range time-varying magnetic fields (much easier to implement) and allowing to recycle the exploited quantum probes (granting shorter acquisition times and increased precision).

\end{abstract}
%\pacs{42.50.-p, 42.50.Ar, 42.50.Dv}
\maketitle

\section{Introduction}
The unification of quantum theory and general relativity is the ``holy grail'' of {contemporary physics}.
Nonetheless, the lack of experimental evidence forbids to discriminate among the different mathematical models \cite{K14,R00,Ved2022}, from string theory \cite{S08} to loop quantum gravity \cite{R08} and supergravity \cite{Cas2022}, or even to establish whether gravity needs to be quantized or if, vice versa, quantum mechanics must be ``gravitized'' at some scale \cite{Opp23}.
Indeed, it has been argued that quantum mechanics might cease to apply at a certain scale, for instance due to the collapse of macroscopic quantum superpositions \cite{A03} possibly caused by gravity \cite{D09,P14}.
Thus, it is crucial to start searching for quantum gravity effects at small scales \cite{Pra2020,Add2022}. %, e.g. along the line of a recent proposal suggesting that the observation of gravity-induced entanglement could demonstrate (or challenge) the quantum nature of gravity.
Notably, Marletto \& Vedral \cite{M17,Mar2017} and Bose et al. \cite{Bos2017} proposed a test of quantum gravitational features- known with the acronym BMW, based on the ability of gravity to generate entanglement between two spatially-superposed massive quantum probes.

The exact meaning of this observation is debated \cite{Add2022,Hug2022}; e.g., in the constructor theory framework \cite{con,Mar22} it witnesses non-classicality of gravity when assuming locality and information interoperability (i.e., that classical information can be copied) \cite{CV12}.
This test could probe quantum effects in gravity in a low-energy regime, largely below Planck's scale.
Moreover, detecting gravitational entanglement would refute a large set of classical gravity theories in presence of quantum masses, e.g. all semiclassical theories as quantum field theory in curved space-time \cite{Bir}, collapse models \cite{A03,D09,P14} predicting collapse at the experiment energy scale, and other hybrid quantum-classical theories \cite{Bar,She}.
Detecting entanglement would therefore represent a major breakthrough, being the first refutation of classical theories of gravity and confirming that gravity must be quantized – thus achieving the first experimental confirmation of quantum effects in gravity.
Conversely, not detecting entanglement would not demonstrate that gravitational interaction is classical, but would still open several interesting possibilities \cite{Bas2017,Add2022} (e.g., one of the assumptions is wrong, other forces are at play, the magnitude of the effect is too small, etc.).

The current challenge lies in realizing this experiment with available technology \cite{Rij2021,How2021,Mar2021,Hen2022,Mar2022,Jap2022,Hen2023,Fen2023,Han2023,Bra2023,Tor2024,Wu2024,Xia2024,Mar2024,Schu2024,Bos2025}.
%Thanks to recent advances in quantum metrology and quantum sensing, one is able to manipulate quantum objects of increasing mass (ranging from nanoparticles to MEMS oscillators), so this task can be deemed to be within reach.
Bose et al. proposed to exploit two nanodiamonds (NDs) presenting single nitrogen-vacancy (NV) centres.
Both NDs are prepared in a spatially-delocalized quantum superposition by means of a magnetic field gradient \cite{Wan2016,Woo2022a} acting on the NV-centre spin, and subsequently they are left subjected only to reciprocal gravitational interaction by letting them fall freely in vacuum.
%Anyway, such a free-fall scheme presents several extreme technical difficulties: it requires building a magnetic structure $\gtrsim10$ m, with fabricated features of the order of micrometers, and ND control on a scale of $\sim100$ nm, both very hard to achieve.
%Furthermore, it requires for such a large structure high-vacuum and low-temperature conditions, not to mention the realization of the antennas delivering the microwave pulses for the NV (or other color centres with a similar spin level structure \cite{Slurm2014}) spin manipulation, required to implement the DD \cite{Ped2020} needed to increase the NV-centres coherence time.\\
However, this scheme presents significant technical challenges, requiring a magnetic structure longer than $10$ m with features fabricated to micrometric precision, to be maintained in high-vacuum and low-temperature conditions.
Furthermore, a large number ($\gtrsim{10^4}$) of antennas delivering microwave (MW) pulses are needed to achieve quantum control of the spin states of NV centres (or similar color centres \cite{Slurm2014}), crucial for extending the superpositions coherence time \cite{Ped2020}.\\

\begin{figure}[th]
	\centering
	\includegraphics[width=\columnwidth]{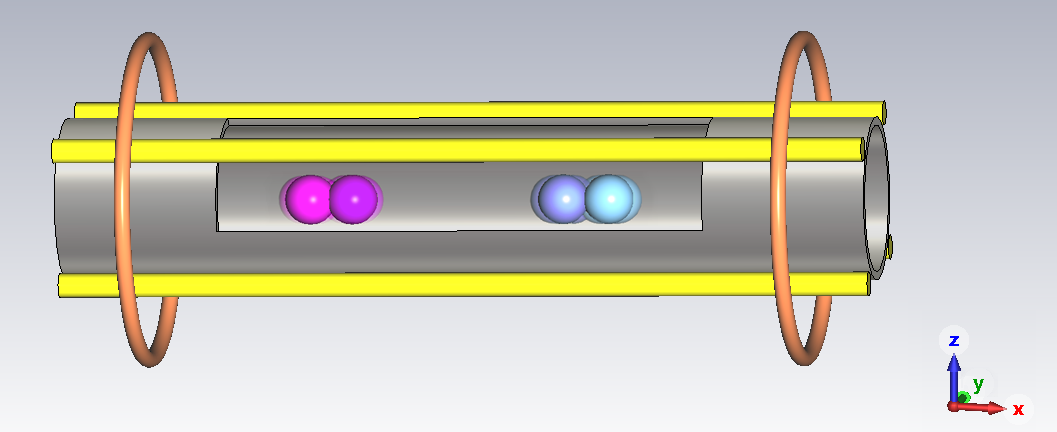}
	%\caption{The physical system under study is a single-NV-centre nanodiamond (trapped in the $y$ and $z$ directions by a gravito-magnetic potential) in a spatially-delocalized superposition, generated by putting the NV center spin component $S_x$ in the initial state $\ket*{\psi_S}=\frac{1}{\sqrt2}(\ket*{-1}+\ket*{1})$ before applying a magnetic field gradient $B'$ along the $x$ direction, thus spatially separating the two components of $\ket*{\psi_S}$. Depending on its $S_x$ direction (indicated by the red arrow), each component will then oscillate in the $x$ direction around a different equilibrium position $x_0^{(\pm)}$, according to Eq. \eqref{motion}.}
\caption{The proposed experiment relies on two single-NV-centre NDs (trapped along $y$ and $z$ by a gravito-magnetic potential, but free along $x$) both in a spatially-delocalized superposition. %, generated by putting the NV center spin component $S_x$ in the initial state $\ket*{\psi_S}=\frac{1}{\sqrt2}(\ket*{-1}+\ket*{1})$ before applying a magnetic field gradient $B'$ along the $x$ direction, thus spatially separating the two components of $\ket*{\psi_S}$.
The distance mismatch between the superposition components of the two NDs should allow the gravitational potential to entangle the NDs, an impossible task for local operations and classical communication.}
%Depending on its $S_x$ direction (indicated by the red arrow), each component will then oscillate in the $x$ direction around a different equilibrium position $x_0^{(\pm)}$, according to Eq. \eqref{motion}.}
	\label{scheme}
\end{figure}

\rev{Conversely, along the path of BMV-like proposals involving trapped massive quantum objects \cite{sch2023,sch2024,ela2025} instead of free-falling ones, here we propose a more straightforward table-top scheme (see Fig. \ref{scheme}) involving semi-trapped NDs, constrained in two directions but free to move along the third one; if gravity presents quantum features, the distance mismatch generated between the steady spatially-separated superposition components of the two NDs should allow the gravitational potential to entangle them, an impossible task for local operations and classical communication (LOCC) \cite{Ben1996,Hor2009}.\\
With respect to previously-considered alternative configurations for macroscopic superpositions of levitated massive objects (see, e.g., \cite{Johnsson2016,Pino2018,Hen2023}), exploiting NDs in optical, Paul or magneto-optical traps \cite{Russell2021,delord2019spin,Delord2020,skakunenko2025strongconfinementnanodiamondneedle,Hen2023, Hodges2026}, diamagnetic levitation with chips or permanent magnets \cite{ela2025,Hsu2016, OBrien2019,Leng2021}, ferromagnets levitation on superconductors \cite{Bernstein2020,Vinante2022,Latorre2022}, our proposal involves a ND magnetostatic trap offering several advantages in gravity-based experiments. 
Besides the fact that in-trap experiments guarantee faster execution, as they make possible to retain the samples after each run, eliminating the need for repeated trapping and characterization steps (heavily slowing down the procedure), our design offers tunable longitudinal control via magnetic field gradients produced by anti-Helmholtz coils, and it allows the ND wavepacket to be expanded to macroscopic scales and subsequently kept in an inertial, delocalized state.
Up to our knowledge, this coil-based design allows, via a single MW antenna and high-frequency AC control, for the first efficient implementation of spin and motional dynamical decoupling (DD) \cite{Ped2020} on trapped objects (as similarly introduced for optical levitation \cite{skrabulis2026}).\\
Interestingly, we observe that in this configuration, the NV centres accumulate no bias-magnetic-field-dependent Zeeman phase (eventually recovered by tilting the setup to introduce a difference in gravitational potential energy), inherently providing protection against fluctuations in the magnetic field and contributing to a reduction in dephasing factors while still maintaining the bias necessary to prevent Majorana spin-flips \cite{Majorana1932}.}

\section{Single nanodiamond interferometer}

For simplicity, let us consider the case in which a stationary single-NV-centre ND is left free to move along $x$ while being constrained in $y$ and $z$, with the NV-centre spin axis aligned along $x$.
Such a scenario can be obtained by exploiting a magnetic trap \cite{Pri1983,Tre2004,For2007,Bau2020}, with the magnetic field forming a 2D confining potential in $y$ and $z$ while leading the dynamics along $x$, possibly hosting doughnut-shaped optical tweezers \cite{Gei13} to safely set the position of the trapped ND in the $x$-$y$ plane. % because of excessive heat.
Initialized the $S_x$ component of the ND NV-centre in the state $\ket*{\psi_S}=\frac{1}{\sqrt2}(\ket*{-1}+\ket*{+1})$, to create a spatially-delocalized ND superposition we generate a magnetic field gradient $B'=\frac{dB}{dx}$.
\rev{The ND Hamiltonian along $x$ is:
\begin{align}
H_x=&\frac{p^2_x}{2m} + \frac{\chi V(B_0 + B'x)^2}{2\mu_0} + \hbar\gamma_eS_x(B_0 + B'x)\nonumber\\
&+ \hbar D S^2_x + mgx\sin\theta_g,
\label{ham}
\end{align}
%\begin{equation}
%H_x=\frac{p^2_x}{2m} + \frac{\chi V(B_0 + B'x)^2}{2\mu_0} + \hbar\gamma_eS_x(B_0 + B'x) + \hbar D S^2_x + mgx\sin\theta_g,
%\label{ham}
%\end{equation}
%
}
being, respectively: $m$, the ND mass; $p_x$, its momentum along $x$; $V$, its volume; $\chi$, its magnetic susceptibility (assuming for the ND a diamagnetic behavior); $\mu_0$, the vacuum magnetic permeability; $\gamma_e$, the gyromagnetic ratio of the electron; $B_0$, the bias magnetic field; $D$, the zero-field splitting of the NV centre\rev{; $g$: gravitational acceleration; $\theta_g$: angle between the $x$ axis and the horizon.}
Classically, the dynamics generated by this Hamiltonian corresponds, for two separated particles of opposite spin, to a harmonic oscillation with identical frequency but different equilibrium positions:
\begin{equation}
x^{(\pm)}(t)= x_0^{(\pm)}(1-\cos(\omega t)),
\label{motion}
\end{equation}
%
%being $x_0^{(\pm)}=-\frac{\chi V B_0\pm\hbar\gamma_e\mu_0}{\chi V B'}$ the equilibrium position of the particle with $S_x=\pm1$ spin component, and $\omega=B'\sqrt{\frac{\chi V}{\mu_0m}}$ the oscillation frequency.
\rev{being $x_0^{(\pm)}=-\frac{\chi V B_0B'+\mu_0mg\sin\theta_g\pm\hbar\gamma_e\mu_0B'}{\chi V (B')^2}$ the shifted equilibrium position relative to the  $S_x=\pm1$ spin components, and $\omega=B'\sqrt{\frac{\chi V}{\mu_0m}}$ the oscillation frequency.
Under appropriate cooling of the center of mass, the spin coupling with quantized phononic oscillations of the center of mass induce a coherent superposition of these trajectories rather than a statistical mixture.
The coherence length $\Delta x(t)=|x^{(+)}(t)-x^{(-)}(t)|$ of the wavepacket oscillates in time:
\begin{equation}
\Delta x(t)=\Delta x_\text{max}\sin^2\left(\frac{\omega t}{2}\right)\;,\;\;\;\;\;\;\Delta x_\text{max}=\frac{4\hbar\gamma_e\mu_0}{\chi VB'}\;,
\label{DeltaX}
\end{equation}
where $\Delta x_\text{max}$ indicates the maximum separation achievable between the two components. Eq. (\ref{DeltaX}) indicates that, in general, to larger samples correspond smaller delocalizations. Also, higher magnetic gradients introduce stronger restoring forces, increasing the oscillation frequency but limiting the wavepacket expansion. }\\
\rev{The quantum mechanical version in terms of phononic normal modes reads:
\begin{equation}
%\hat{H}_x= \hbar\omega \hat{a}^\dagger \hat{a} + \hbar\lambda(\hat{a}+\hat{a}^\dagger)\hat{S}_x + \hbar\gamma_e B_0\hat S_x + \hbar D \hat{S}^2_x,
\hat{H} = \hbar\omega \had\ha + \hbar\left(\lambda_0+\lambda \hS\right)(\ha+\had) + \hbar\gamma_e B_0\hS+\hbar D \hS^2\;,
\label{hamq}
\end{equation}
%
%with $\lambda=x_\text{zpf}\gamma_e B'$ and canonical bosonic commutation relations $[\ha,\had]=1$ relating to the position and momentum by $\hx=x_\text{zpf}(\ha+\had)$ and $\hp=ip_\text{zpf}(\had-\ha)$, where we have introduce the zero-point standard deviations $x_\text{zpf}=(\hbar/2m\omega)^{1/2}$ and $p_\text{zpf}=(\hbar m \omega/2)^{1/2}$. }
with $\lambda_0 = ( \chi V B_0B' / \mu_0 +mg\sin\theta_g)x_\text{zpf}/\hbar $ and $\lambda=\gamma_e B' x_\text{zpf}$ respectively, and canonical bosonic commutation relations $[\ha,\had]=1$ relating to the position and momentum by $\hx=x_\text{zpf}(\ha+\had)$ and $\hp=ip_\text{zpf}(\had-\ha)$, where we have introduce the zero-point standard deviations $x_\text{zpf}=(\hbar/2m\omega)^{1/2}$ and $p_\text{zpf}=(\hbar m \omega/2)^{1/2}$.}
%Conversely, the quantum mechanical version reads:
%
%\begin{align}
%H_x&=\frac{\hat{p}^2_x}{2m} + \frac{\chi V(B'\hat{\xi})^2}{2\mu_0} + \hbar\gamma_e\hat{S}_x(B'\hat{\xi}) + \hbar D \hat{S}^2_x \nonumber\\
%&= \hbar\omega \hat{a}^\dagger \hat{a} + \hbar\lambda(\hat{a}+\hat{a}^\dagger)\hat{S}_x + \hbar D \hat{S}^2_x,
%\label{hamq}
%\end{align}
%
%with $\hat{\xi}=\hat{x}+\frac{B_0}{B'}$ and $\lambda=\gamma_e B'\sqrt{\frac{\hbar}{2m\omega}}$.
The spatial component of the eigenstates for $S_x=\pm1$ describes a shifted harmonic oscillator with ground states:
\begin{equation}
\Big|{\frac{\lambda_\pm}{\omega}}\Big\rangle=e^{-\frac12\cdot\left|\frac{\lambda_\pm}{\omega}\right|^2}\sum_{n=0}^\infty\left(\frac{\lambda_\pm}{\omega}\right)^n\frac{(\hat{a}^\dagger)^n}{n!}\ket{0}\;,\\
%S_x=\pm1:\;\;\;\;\ket*{\frac{\lambda_\pm}{\omega}}=e^{-\frac12\cdot\left|\frac{\lambda_\pm}{\omega}\right|^2}e^{\frac{\lambda_\pm a^\dagger}{\omega}}\ket*{0}\;,
\label{ground}
\end{equation}
being $\lambda_\pm=\lambda_0\pm\lambda$, corresponding to coherent states with amplitude $\alpha=\lambda_\pm/\omega$ (see Appendix \ref{Appendix: Dynamics}).
%Applying the evolution operator $\hat{U}=\exp(-\frac{i}{\hbar}\hat{H}_xt)$ to our ND, one obtains:
%
%\begin{equation}
%\hat{U}\Big[\ket{\psi_S}\otimes\ket0\Big]=\frac{ \ket{\alpha(t)}_+\ket{+1}+ e^{i\Phi(t)~}\ket{\alpha_-(t)}\otimes\ket{-1}}{\sqrt2}\;,
%\label{evol}
%\end{equation}
%
%\begin{align}
%\ket{\alpha(t)}_\pm =& e^{i\left[\left(\left|\frac{\lambda_\pm}{\omega}\right|^2-\frac12\right)\omega t+\left|\frac{\lambda_\pm}{\omega}\right|^2\sin(\omega t)\right]}\times\nonumber\\
%&\times\ket{\frac{\lambda_\pm}{\omega}\left(1-e^{i\omega t}\right)}\;.
%\label{evol}
%\end{align}
%which indicates the overall state is a superposition of two coherent states for each spin component, with amplitude 
%
%\begin{equation}
%%\ket{\alpha(t)}_\pm = e^{i\left[\left(\left|\frac{\lambda_\pm}{\omega}\right|^2-\frac12\right)\omega t+\left|\frac{\lambda_\pm}{\omega}\right|^2\sin(\omega t)\right]}\ket{\frac{\lambda_\pm}{\omega}\left(1-e^{i\omega t}\right)}\;.
%\alpha_\pm(t) = \frac{\lambda_\pm}{\omega}\left(e^{i\omega t}-1\right)
%\label{evol2}
%\end{equation}
%
%and relative time-dependent phase 
%
%\begin{equation}
%\Phi(t)= -\frac{(\lambda_+-\lambda_-)^2}{\omega^2}\big( \omega t+\sin(\omega t)\big) +2\gamma_eB_0t
%\label{theta_pm}
%\end{equation}
%
%%with $\theta_\pm(t)=\left[\left(\left|\frac{\lambda_\pm}{\omega}\right|^2-\frac12\right)\omega t+\left|\frac{\lambda_\pm}{\omega}\right|^2\sin(\omega t)\right]$.
Applying the evolution operator $\hat{U}=\exp(-\frac{i}{\hbar}\hat{H}t)$ to our ND, initially considered in the harmonic oscillator ground state $\ket*{0}$, one obtains \rev{(up to a global phase):
%
%\begin{equation}
%\hat{U}\big(\ket{\psi_S}\ket0\big)=\frac{\ket{-1}\ket{\alpha(t)}_- + \ket{+1}\ket{\alpha(t)}_+}{\sqrt2}\;,
%\label{evol}
%\end{equation}
%
\begin{equation}
\hat{U}\big(\ket{\psi_S}\otimes\ket0\big)=\frac{\ket{-1}\otimes\ket{\alpha_-(t)} +e^{i\Delta\theta(t)}\ket{+1}\otimes\ket{\alpha_+(t)}}{\sqrt2}\;,
\label{evol}
\end{equation}
with coherent amplitudes
\begin{equation}
\alpha_\pm(t) = \frac{\lambda_\pm}{\omega}\left(e^{i\omega t}-1\right)\;
\label{evol2}
\end{equation}
and relative phase $\Delta\theta(t)=\theta_+(t)-\theta_-(t)$, being
\begin{equation}
\theta_\pm(t)=\left(D\pm\gamma_eB_0-\frac{\lambda_\pm^2}{\omega}\right)t+\left(\frac{\lambda_\pm}{\omega}\right)^2\sin(\omega t)\;.
\label{theta_pm}
\end{equation}
}\\
For $B_0\rightarrow0$, Eq. \eqref{evol} describes for the two $S_x$ components an oscillatory dynamics with equilibrium positions symmetric with respect to the ND rest position (i.e. the ``0'' of the $x$ axis); the bias magnetic field $B_0$ shifts concurrently the equilibrium position of both the harmonic oscillations.
\begin{figure}[th]
	\centering
	\includegraphics[width=\columnwidth]{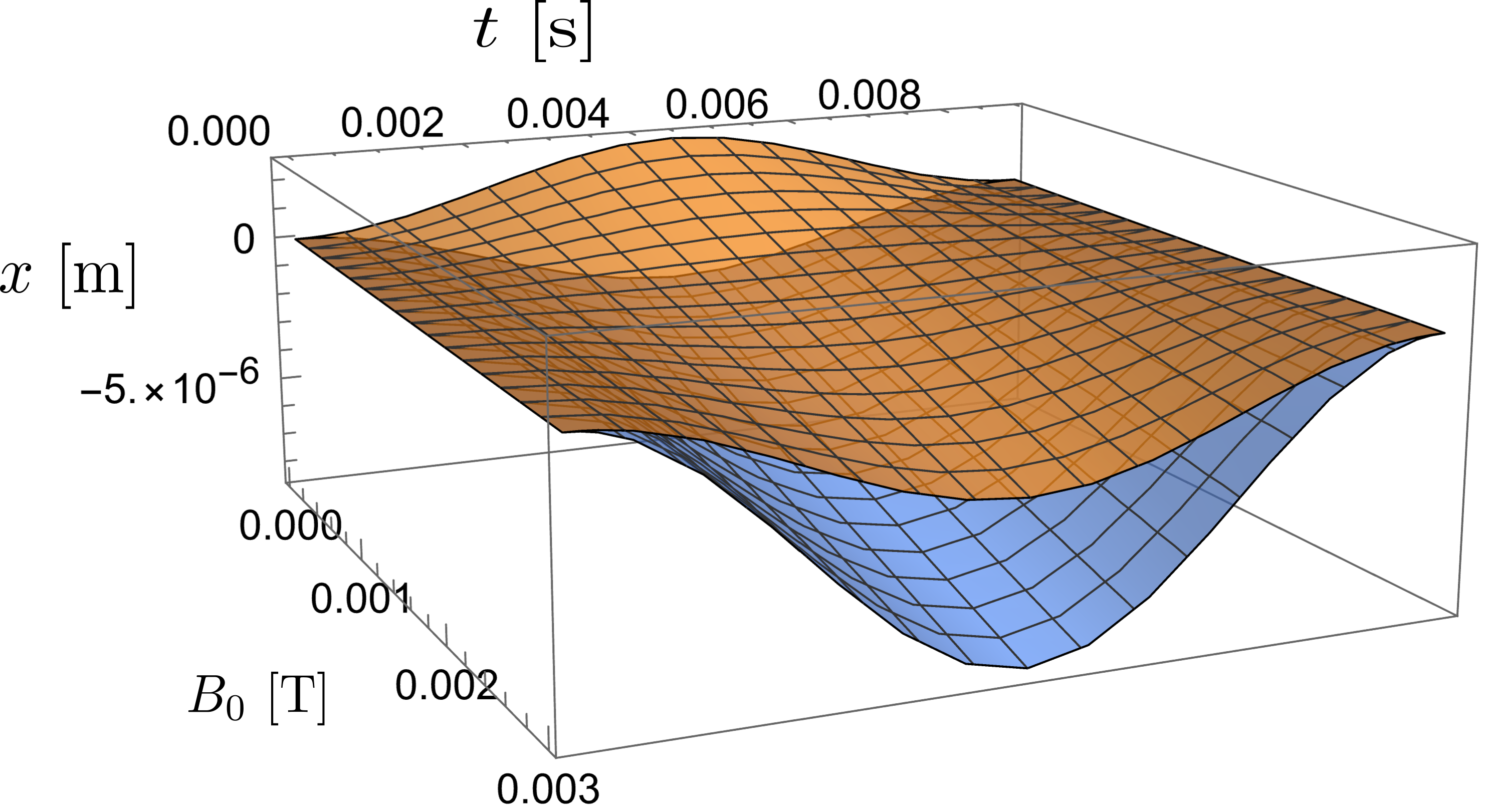}
	\caption{One-period motion along $x$ of the $S_x=+1$ (orange curve) and $S_x=-1$ (blue curve) components of a 250 nm diameter ND in a magnetic field gradient $B'=10^3$ T/m as a function of the bias magnetic field $B_0$.%, shifting of the same amount the equilibrium position of both superposition components.
}
	\label{mm2.9}
\end{figure}
In Fig. \ref{mm2.9} we plotted a single oscillation trajectory for both the $S_x=\pm1$ components, considering a ND with a 250 nm diameter, a magnetic field gradient (along the $x$ axis) $B'=10^3$ T/m and a bias magnetic field $B_0$ varying from 0 to 3 mT \rev{($\theta_g=0$)}.
This value for the ND diameter was chosen as a case study for the table-top interferometer to allow for a direct comparison with previous literature (see, e.g., Ref. \cite{Woo2022a}), although in principle NDs of this size are probably too small for observing some gravity-induced entanglement.
Nevertheless, given their generality, the validity of the obtained results remains the same (a systematic analysis of the gravity-induced entanglement for different ND masses will be shown in Section III ).
For $B_0=0$, the ND superposition components follow symmetric trajectories, while by increasing $B_0$ the equilibrium positions of both oscillatory motions are shifted of the same amount, leaving the maximum separation between them unchanged.
\rev{Remarkably, a full oscillation around the new equilibrium positions accumulates no relative phase depending on $B_0$. This implies that an interferometer such built is naturally insensitive to fluctuations of homogeneous magnetic fields at recombination, shielding its performances from the limitations introduced by uncontrolled biases.
This occurs because the spin-dependent force creates new local equilibrium positions within the inhomogeneous magnetic gradient, and the additional phase accumulated because of this displacement exactly compensates the Zeeman phase (see Appendix \ref{Appendix: Dynamics}).
%This effect can be interpreted as a consequence of diamond diamagnetic moment, which leads to equilibrium positions at the minima of the magnetic field. This is the reason that requires a tilt of the experiment, to accumulate a difference of phase given by the difference in the gravitational potential.
To accumulate a non-zero $\Delta\theta$, one can tilt the setup of a (tiny) angle $\theta_g$, inducing a difference in the gravitational potential between the two superposition components and yielding, at recombination time $T=2\pi/\omega$:
%At recombination time $T=2\pi/\omega$ the relative phase accumulated is:
%
\begin{equation}
    \Delta\theta(T) = -4\pi\left(\frac{\mu_0 m}{\chi V} \right)^\frac32\frac{\gamma_eg}{B'^2}\sin\theta_g\;.
\end{equation}
%For a magnetic gradient of $B'=100$ T/m corresponding to $T=0.87$ s, a tilt of an angle $\theta_g\approx 10^{-9}$ rad produce a massive phase difference of $\Delta\theta \approx 2*10^7$ rad.
%Small run-to-run variation of the angle would produce great phase oscillations, averaging out the interference pattern, and requiring extremely careful initial calibration. However, the accumulated phase depend on the \textit{relative difference} of altitudes between the superimposed paths, meaning the the only source of systematic errors can be introduced by vibrational noise of the setup or undetected small drifts of the tilting angle.

%
The coherences between the electronic spins are probed using a Ramsey-type sequence that extracts these coherences from the population differences of the excited energy levels.
This will be done introducing a final $\pi$-pulse selective on the $\ket{+1}\rightarrow\ket{0}$ transition to transfer the phase difference $\Delta\theta$ into an optical contrast between $\ket{0}$ and $\ket{-1}$.
This protocol enables to test the authenticity of the ND macroscopic superposition state, allowing to explore the range of feasibility of quantum gravity experiments with large masses. 
It must be noted that the macroscopicity of these superpositions, to be intended in terms of mass, approaches the state of the art for the observation of matter-wave phenomena with large objects \cite{Pedalino2026, Bateman2014,RomeroIsart2010}.
Furthermore, as the phase difference depends only on the gravitational phase, this platform serves as a natural testbed for gravimetry \cite{z8b4-sm79, Johnsson2016,headley2026quantummetrologynewtonsconstant,Mnoret2018}.
}

%-----Remove---------------------------------------------Simple algebra shows that, for a complete oscillation (i.e., $t_1=2\pi/\omega$), one has $\theta_+(2\pi/\omega)=\theta_-(2\pi/\omega)$, therefore no phase difference is generated between the two superposition components.
%\vspace{0.5cm}
%To check the coherence of our superposition as a phase difference accumulated in a Ramsey-like interferometric measurement \cite{Van2004}, we can tilt the interferometer of a (small) angle $\theta_g$ in the $x-z$ plane, generating a gravitational potential mismatch between the two ND superposition components due to the different position in $z$ \cite{sca2013}.
%This reflects in an additional $g$-depending term appearing in the Hamiltonian of Eq. \eqref{ham}, being $g$ the gravitational acceleration, resulting in a phase difference (after a complete oscillation):!!!
%%
%\begin{equation}
%\Delta\vartheta=\vartheta_+(t_1)-\vartheta_-(t_1)= -4\pi\left(\frac{\mu_0m}{\chi V} \right)^{3/2}\frac{\gamma_eg}{B'^2}\sin\theta_g\;.
%\label{delta_Theta}
%\end{equation}
%%
%By repeating this procedure several times with different $\theta_g$ values, one can characterize the ND spatial superposition, determining if its coherence allows for quantum gravity experiments.----------------------------\\

\rev{
\subsection{Decoherence and other instability sources}
We now quantify the realistic performance of the interferometer counting in open-system environmental effects.
The interaction with uncontrolled external factors will, in general, have detrimental consequences on the overall visibility. The contributions can be divided into three main areas: spin decoherence, motional decoherence, systemic dephasing and recombination errors. \\
The decoherence sources for NV centers in NDs can be divided into bulk- and surface-related ones.
The latter are due to paramagnetic dangling bonds and strain, while the former ones are associated with the presence of fluctuating spin baths of $^{13}\text{C}$ and P1 centers (in which a single nitrogen atom replaces a carbon atom in the carbon grid \cite{P1a,P1b}).
%The fluctuating field due to dangling bonds scales as $r^{-3}$, being $r$ the distance of the NV center from the ND surface, and is generally negligible for $r\gtrsim250$ nm compared to the bulk $^{13}\text{C}$ bath \cite{janitz2022diamond,Knowles2013}.
The variance of the fluctuating field due to the ND surface dangling bonds scales as $r^{-4}$, being $r$ the distance of the NV center from the ND surface, and for $r\gtrsim250$ nm it is generally negligible with respect to the contribution from bulk $^{13}\text{C}$ bath \cite{janitz2022diamond,Knowles2013}.
Regarding strain, the fabrication process introduces a strain gradient in the ND, and variations in this strain arising from thermal or mechanical vibrations can lead to decoherence.
If care is taken in the fabrication of the microstructure, the decoherence due to strain can be mitigated \cite{And14}.
For microstructures with dimensions around 1 $\mu\text{m}$ (like the ones suggested for the gravity experiment in Section III), the surface sources of decoherence are strongly mitigated.
Hence, only the decoherence coming from $^{13}\text{C}$ and P1 baths has to be taken into account, and we can consider $T_2\sim10$ s with DD techniques, as demonstrated in a recent work \cite{yamamoto2026ten}. 

\noindent
Motional decoherence of the ND center of mass is, to leading order, determined by scattering processes involving blackbody radiation and residual gas particles.
Elastic collisions with background photons or air molecules reveal which-path information by entangling it with environmental degrees of freedom, thereby destroying the superposition.
The nature of this effect depends strongly on the wavelength of the scattering particles, i.e. its resolving power, and divides in the limit-cases of Short- and Long-Wavelength Limit (SWL and LWL) for $\lambda_\text{env}/\Delta x \ll1$ or $\lambda_\text{env}/\Delta x \gg1$ respectively \cite{Schlosshauer2019,PhysRevA.111.042211}.\\
Air molecules (predominantly nitrogen with mass $m_g = 4.65\times10^{-26}$ kg) possess a thermal de Broglie wavelength $\lambda_\mathrm{th}=\frac{2\pi\hbar}{\sqrt{2\pi m_gk_BT_\mathrm{e}}}\simeq 0.1$ nm at the environmental temperature $T_e=4$ K ($k_B$: Boltzmann constant).
For the masses and magnetic gradients under consideration, while the maximal separation $\Delta x_\text{max}$ in Eq. \eqref{DeltaX} always results deep in the SWL, the spread close to opening and recombination phases is $\lambda_\text{th}/x_\text{zpf}\sim 1$, requiring interpolation in between SWL and LWL to precisely estimate collisional decoherence.
A quantitative estimation can be obtained from the Quantum Linear Boltzmann Equation \cite{PhysRevLett.97.060601,PhysRevA.77.022112,Vacchini2009}, which, upon neglecting van der Waals short-range interactions \cite{PhysRevA.111.042211,Vacchini_2004}, yields the decoherence rate
\begin{equation}
    \label{equ:GammaGWL}
    \Gamma_\text{air} = \Gamma_\text{SWL}\left(1+\frac{\text{Da}\left( \xi\right)}{\xi} \right)
\end{equation}
with
\begin{equation}
    \label{equ:UsefulDefs}
    \Gamma_\text{SWL} = \frac{4\pi \hbar n_g R^2}{m_g \lambda_\mathrm{th}}~~~, \quad \xi=2\sqrt{\pi}\frac{\Delta x}{\lambda_\text{th}}~~,
\end{equation}
which correctly retrieves the SWL and LWL in the asymptotic limits of the Dawson function $\text{Da}(X) = \int_0^\infty e^{-t^2}\sin(2Xt)$~\cite{PhysRevA.111.042211} ($n_g$: gas particle number density; $R$: ND radius).
Eq.~\eqref{equ:GammaGWL} saturates to the SWL decoherence rate given in Eq. \eqref{equ:UsefulDefs}, which depends on the gas particle number density $n_g = P/k_B T_e$  ($P$: pressure), taken for an ideal gas.\\
Decoherence arising from thermal radiation encompasses elastic Rayleigh scattering of ambient blackbody photons, alongside inelastic absorption and emission processes.
The Planck spectrum peaks at the blackbody wavelength $\lambda_\text{BB} = \pi^{2/3}\frac{\hbar c}{k_BT_\text{e}}$ (being $c$ the speed of light in vacuum), which results in the microwave/terahertz regime $\lambda_{\text{th}} \sim 1\;\text{mm}$ at $T_\text{e}=4$ K.
In the parameters space under analysis, the wavepacket separation always results much smaller; this implies blackbody decoherence is described in the LWL limit by the rate 
\begin{equation}
    \Gamma_\text{BB}= \Lambda_\text{BB}(\Delta x)^2~~~,
\end{equation}
scaling quadratically with the wavepacket separation.
The localization parameter $\Lambda_\text{BB}=\Lambda_\text{sc}+\Lambda_\text{a}+\Lambda_\text{e}$ includes contributions from scattering, absorption and emission processes \cite{PhysRevA.84.052121}:
\begin{align}
    \Lambda_\text{s} &= \frac{8c\times8!~\zeta(9)}{9\pi}\Bigg|\frac{\epsilon_\text{bb}-1}{\epsilon_\text{bb}+2}\Bigg|^2 \left(\frac{k_BT_e}{\hbar c}\right)^9R^6~~,\\
    \Lambda_\text{a(e)} &= \frac{2c\times5!~\zeta(6)}{3\pi}~\mathcal{I}\left[\frac{\epsilon_\text{bb}-1}{\epsilon_\text{bb}+2}\right] \left(\frac{k_BT_{e(i)}}{\hbar c}\right)^6R^3~~,
\end{align}
being $\zeta(X)=\sum_{n=1}^\infty n^{-X}$ the Riemann Zeta function and $\epsilon_{bb}$ the ND bulk relative permittivity. 
The localization rates are obtained modeling the ND as a dielectric sphere with $\epsilon_\text{bb} \simeq 5.7$, which exhibits negligible dispersion across the relevant thermal band \cite{PhysRevA.105.032411}.
We assume a conservative upper-bound for the imaginary permittivity of $\Im[\frac{\varepsilon_\text{bb}-1}{\varepsilon_\text{bb}+2}] \approx 10^{-3}$ (extrapolated from high-temperature optically-levitated NDs \cite{Frangeskou2018}); however, we note that the intrinsic cryogenic dielectric loss of diamond is likely orders of magnitude lower.
The rate of blackbody emission is strictly governed by the internal temperature $T_i$ of the ND.
Because diamond possesses an extremely high Debye temperature ($T_D \approx 2200\text{ K}$), the absorption of low-energy environmental photons into the phonon bath is strongly suppressed, allowing $T_i$ to remain decoupled from the vacuum chamber.\\
The relative contributions of the different decoherence mechanisms and their impact on the interferometric visibility are illustrated in Fig. \ref{DecoTot}.
\begin{figure}[htbp]
    \centering
    \includegraphics[width=0.45\textwidth]{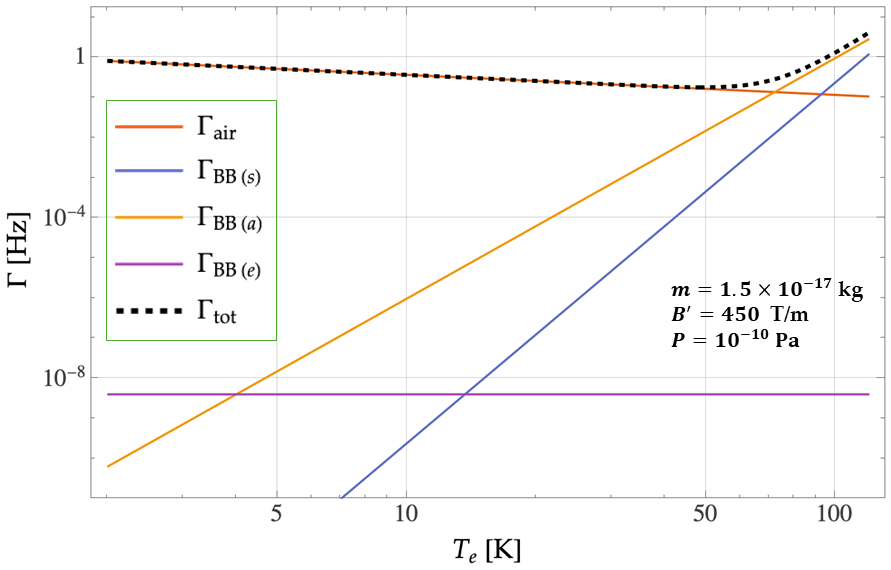}\\
    \includegraphics[width=0.45\textwidth]{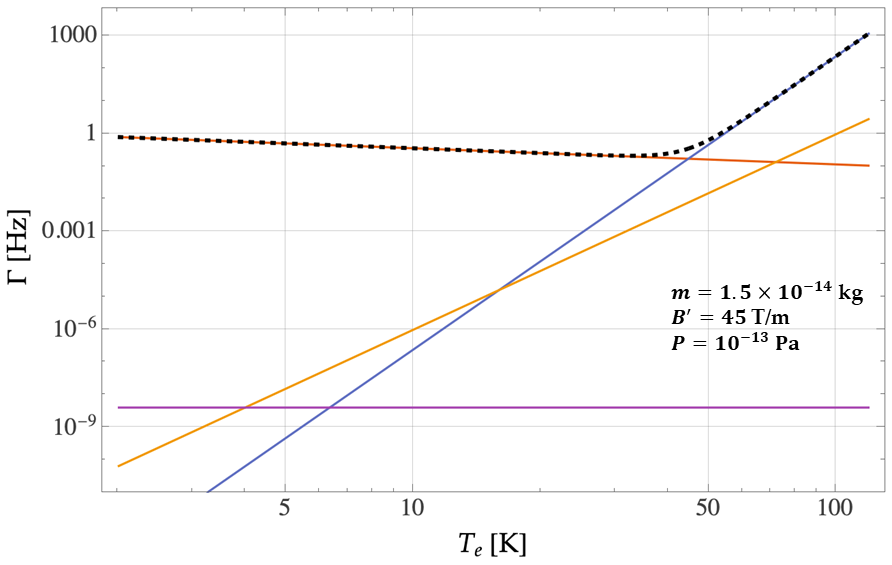}
    \caption{\rev{Contributions to the total decoherence rates for different ND mass and magnetic field gradient values.
    Decoherence due to scattering with (residual) air molecules scattering decoherence dominates for low temperatures, while blackbody contributions become relevant close to room temperature due to low-optical activity of the crystalline structure.
    Emission is governed by internal temperature, taken constant across the spectrum at $T_i =4$ K.}}
    \label{DecoTot}
\end{figure}
We examine here two representative ND samples with $m = 1.5 \times 10^{-17}$ kg %($\sim 10^{10}$ amu) 
and $m = 1.5 \times 10^{-14}$ kg% ($\sim 10^{13}$ amu)
, respectively.
At low temperatures, decoherence due to residual air molecules dominates by several orders of magnitude, being the primary source of coherence loss.
Blackbody–induced decoherence becomes relevant only near room temperature, owing to the strong thermal isolation of the diamond lattice and its low optical absorption coefficient at cryogenic temperatures.
In general, larger $m$ corresponds to smaller achievable spatial delocalizations, which can be compensated by using weaker $B'$ values.
However, weaker $B'$ values imply lower oscillation frequencies and, therefore, longer times over which noise can accumulate, ultimately leading to greater loss of coherence, as it can be observed in the interferometric visibility plots reported in Fig. \ref{VisTot}.
\begin{figure}[htbp]
    \centering
    \includegraphics[width=0.45\textwidth]{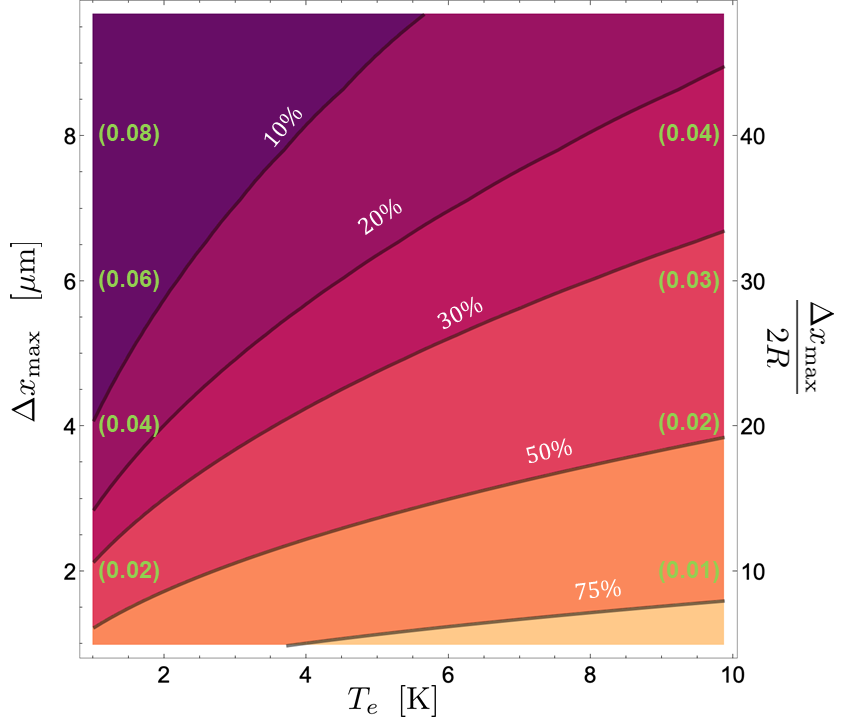}
    \caption{\rev{Interferometric visibility vs. temperature for two sample masses, i.e. $10^{-17}$ kg ($R=0.1~\mu$m, black axes) and $10^{-14}$ kg ($R=1~\mu$m, green axes), at pressure $P=10^{-10}$ Pa and $10^{-13}$ Pa, respectively. Right-vertical axis indicates delocalization size compared to the sample diameter.}}
    \label{VisTot}
\end{figure}
This means that working with increasingly massive systems imposes more stringent experimental requirements for observing interference fringes, potentially fulfilled by operating under Extreme-High Vacuum (XHV) conditions \cite{XHV}.\\
Finally, premature or delayed recombination pulses leave the internal spin states entangled with the spatial wavepackets; this residual which-path information acts equivalently to an additional decoherence channel.
Consequently, the optimal readout timing occurs exactly when the diamagnetic restoring force recombines the center-of-mass wavepackets at the origin, i.e., after one complete mechanical oscillation.
This stringent timing condition becomes evident when computing the interferometric visibility, which is governed by the overlap of the coherent states:
\begin{equation}
\left| \bra{\alpha_-}\alpha_+\rangle \right| \approx \exp\left[ - \left(\bar{n}+\frac{1}{2}\right)\left(\frac{\Delta x_\text{max}}{2x_\text{zpf}} \right)^2\sin^2\left(\frac{\omega\Delta t}{2}\right) \right]~~,\end{equation}
which incorporates the finite thermal phonon occupation $\bar{n}$ of the center-of-mass mode, prepared close to the ground state ($\bar{n}\in [10^{-1},10]$) via standard feedback cooling techniques \cite{FC1,FC2,FC3}. Note that the approximation sign is because with a non-zero initial thermal phononic occupation a pure state picture is not adequate. The generalization of the overlap can be easily obtained via Gaussian state theory of quasiprobability distributions as the Wigner Characteristic or the Glauber-Sudarshan functions \cite{ferraro2005gaussianstatescontinuousvariable,Walls2025}, resulting in the right hand side expression. The loss of visibility as a function of a recombination timing error $\Delta t$ is shown in Fig. \ref{DecoMism}.
In general, microwave control of the NV electronic spin relies on fast Rabi oscillations with frequency $\Omega_R= \gamma_e B_\text{MW}/2$ ($B_\text{MW}$: microwave magnetic field).
Suppressing recombination-induced errors requires control over timescales 
\begin{equation}
    \tau_\text{MW}\ll|\Delta t| ~~~,
\end{equation}
with $\tau_\text{MW}={2\pi/\Omega_R}$, a condition that is satisfied for both samples under laboratory-accessible $B_\text{MW} \approx 10^{-4}\ \text{T}$, yielding timing control at the level of hundreds of nanoseconds.

\noindent 
The main source of systematic errors are patch potentials (creating spatially-varying electric forces), stray magnetic fields, and rotational libration dynamics lying beyond the simple point-particle model, each of them responsible of uncontrolled relative phase shifts degrading the spatially-superposed ND recombination.
Although neutralizing the net charge of the samples effectively suppresses direct Coulomb interactions, the permanent electric dipole of the ND, arising from residual charges, still produces stochastic, run-dependent phase noise.
Assuming a simple fixed-charge-separation picture in which the dipole is generated by a single elementary charge $e$ displaced from its compensating charge by a characteristic distance $d_q = 100$ nm, the corresponding permanent dipole moment is $p_{\text{perm}} = 1.6 \times 10^{-26}~\text{C}\cdot\text{m}$ \cite{PhysRevA.109.L030401, ela2025,Fragolino2024}.
Typical stray surface potentials are of the order of $V_\text{patch}\sim $1–$100$ mV over characteristic patch sizes $\ell_\text{patch}\sim 100$ nm, generating electric fields $E_\text{patch} \approx V_\text{patch}\ell_\text{Patch}/d^2\sim 0.01$–$1$ V/m at distances $d\sim 100~\mu$m from the nearest surface \cite{Gar2020,bul26}.
Hence, the accumulation of dipole-induced potential differences along the interferometric paths at distance of $\Delta x\sim1\;\mu\text{m}$ leads to dephasing factors $\Delta\phi^2\approx(p_\text{perm}\Delta x ~\tau/\hbar)^2\sigma^2_{\nabla E_\text{patch}}\approx 10^8~\text{rad}^2$ for seconds-long interaction times $\tau$ and electric-field-gradient fluctuations $\sigma_{\nabla E_\text{patch}}^2\approx10^4$–$10^6~\text{V}^2/\text{m}^4$. 
While this estimate likely overstates the true effect, its large value nonetheless highlights the necessity for ultra-clean, monocrystalline, or specially engineered noble-metal surface coatings allowing to achieve $V_\text{patch}\ell_\text{patch}\sim 10^{-15}$ V$\cdot$m, so that dephasing remains negligible.\\
Particular attention must be devoted also to magnetic field gradients arising from imperfections in the coils. These contributions can be precisely characterized and compensated for by mapping the mechanical trapping potential via the ND motional Power Spectral Density (PSD) \cite{PhysRevLett.131.043603}, in combination with in-situ magnetic field imaging using the NV center as a local gradient sensor \cite{Rondin2014}. In Sec. II-B, however, we demonstrate that suitably designed microwave pulse sequences can dynamically decouple both the motional and spin degrees of freedom from homogeneous forces, thereby loosening the experimental constraints.\\
Lastly, we account for the fact that different rotational dynamics between the two branches of the superposition lead to an additional loss of visibility \cite{Ped2020}.
Yet, since quantum trapping and control of these vibrational modes have already been achieved \cite{Delord2017}, this effect now constitutes an engineering challenge rather than a fundamental limitation.
\begin{figure}[htbp]
    \centering
    \includegraphics[width=0.45\textwidth]{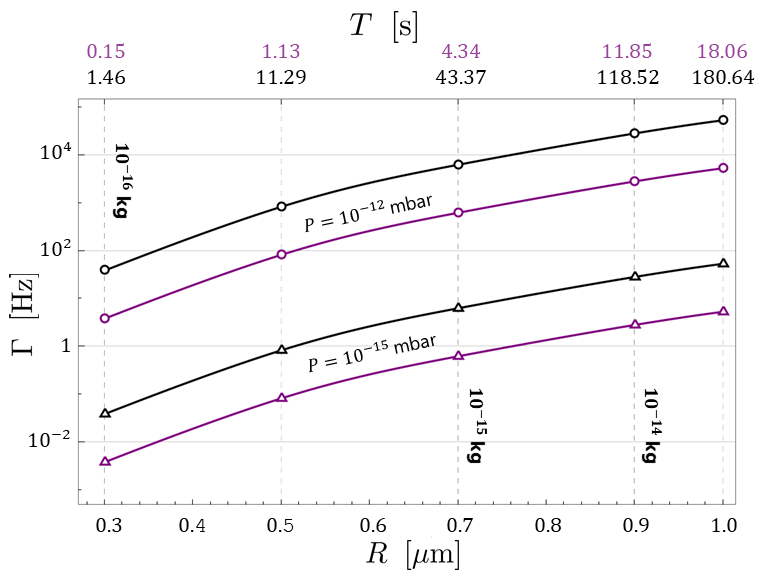}\\
    \includegraphics[width=0.45\textwidth]{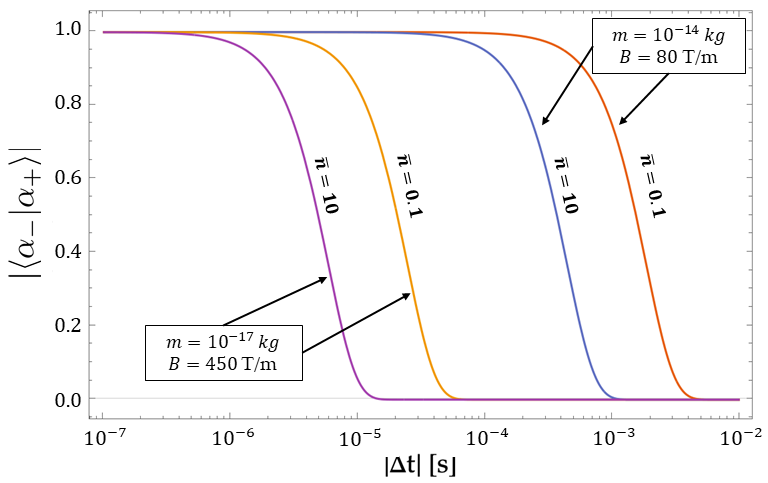}
    \caption{\rev{Top plot: total decoherence evaluated at recombination times as a function of the test mass. Black (purple) curves denote isocontours of the normalized maximum spatial splitting, $\Delta x_\text{max}/2R = 1$ ($10^{-1}$), across various pressure regimes. In the low-temperature limit, residual gas scattering acts as the primary decoherence bottleneck, constraining the realization of macroscopic superposition of larger masses.
    Bottom plot: Visibility reduction due to premature or delayed recombination, indicated by the time mismatch $|\Delta t|$. For initial thermal occupations $\lesssim10$ phonons and stricter requirements of larger $B'$ values, the plot shows that $\mu$s-scale time resolution is sufficient to avoid significant losses.}}
    \label{DecoMism}
\end{figure}
}

%\clearpage

\subsection{Introducing Dynamical Decoupling}

In previous works \cite{Bos2017,Woo2022a} it has been stressed that one possible issue towards the realisation of the quantum gravity experiment is the fact that the time interval needed for generating gravity-induced entanglement between two spatially-delocalized ND superpositions might be $\sim1$ s \cite{Bos2017}.
Actually, this prediction is even optimistic; indeed, in the following we demonstrate that a more realistic scenario requires a time span $\sim150$ s.
Thus, the realization of this experiment will require a significant improvement of the NV-centre coherence time with respect to the present state of the art \cite{Bar2013,Ber2017,Woo2022b}.
In this regard, the NV-centres coherence time can be increased by a DD mechanism \cite{Ped2020}, obtained with a train of microwave $\pi$ pulses flipping the sign of $S_x$ at a frequency $\omega_{DD}=N \omega$.
In order to keep the Hamiltonian in Eq. \eqref{hamq} time invariant, one should flip the direction of $B_0$ and $B'$ (i.e., change both their signs) together with the $S_x$ sign flip.
This way, the system dynamics should remain the same (except for some potential bias due to the small but still non-zero flipping time of $B'$ and $B_0$) given by Eq. \eqref{evol}.\\
%However, having a precise enough control of both $B_0$ and $B'$ to grant such a dynamics could be way too challenging, hence one might consider flipping only $B'$ together with $S_x$, leaving $B_0$ untouched.
To achieve a locally-confined uniform $B'$ and a bias field $B_0$ flipping together in case of DD, an anti-Helmholtz configuration could be chosen (i.e. a Helmholtz coil with two electromagnets carrying currents with opposite directions) \cite{Jan2020}, so that the flipping of $B'$ can be easily driven by supplying the coil windings with radiofrequency currents. %with frequency $\sim100$ kHz.
This way, to flip simultaneously $B_0$ and $B'$ an inversion of the supply currents must be introduced, trying to avoid delays to the ND dynamics which might negatively affect the DD.\\
If the currents have the same amplitude, $B_0\simeq0$ at the target region centre, apart from background noise.
To achieve $B_0\neq0$ along $x$, two solutions can be adopted: (i) the coil windings are supplied with different amplitude currents, producing a dissymmetry in the magnetic field spatial distribution; (ii) the ND is trapped at $x\neq0$, at the cost of a reduced uniformity in $B'$.

\begin{figure}[t]
\begin{subfigure}[c]{0.48\textwidth}
		\centering
		\includegraphics[width=\textwidth]{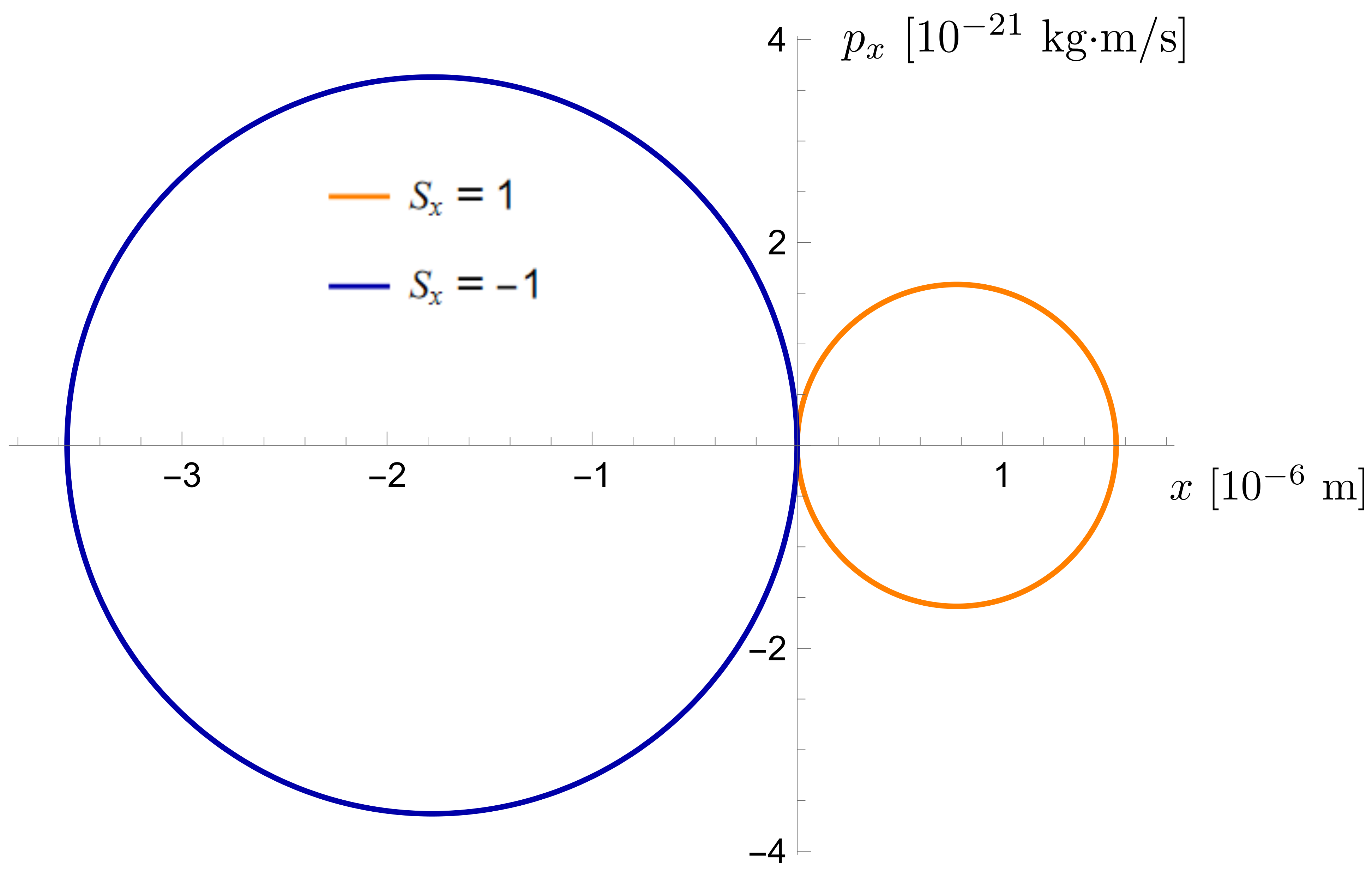}
		\subcaption{$B_0=5\times10^{-4}$ T; no DD.}
		\label{NN0}
	\end{subfigure}
	\hfill
	\begin{subfigure}[c]{0.48\textwidth}
		\centering
		\includegraphics[width=\textwidth]{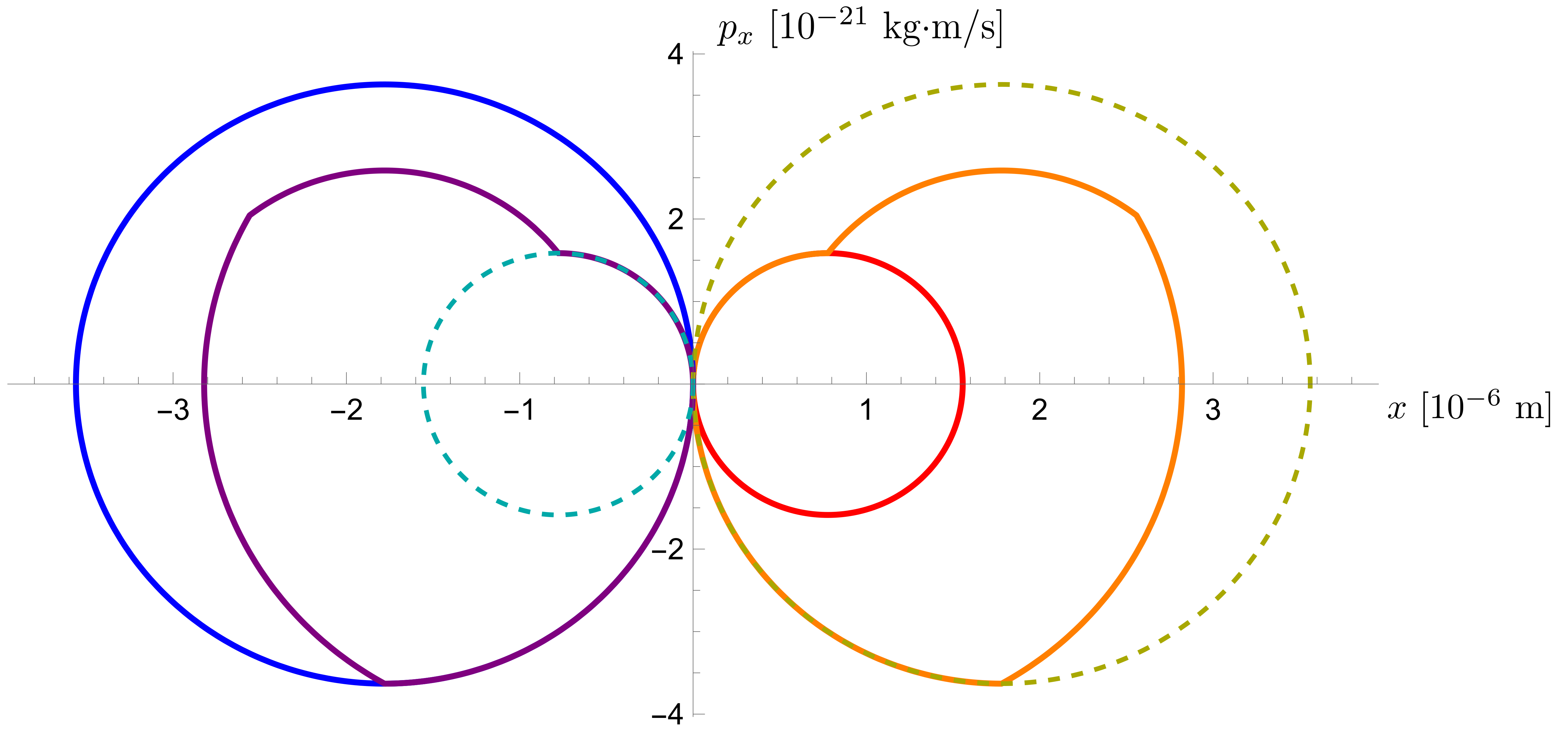}
		\subcaption{$B_0=5\times10^{-4}$ T; $\omega_{DD}=4\omega$.}
		\label{NN4}
	\end{subfigure}
	\hfill
	%\begin{subfigure}[c]{0.48\textwidth}
%		\centering
%		\includegraphics[width=\textwidth]{phase_DD_NN20_B10-3.3new.png}
%		\subcaption{$B_0=5\times10^{-4}$ T; $\omega_{DD}=20\omega$.}
%		\label{NN20}
%	\end{subfigure}
%	\hfill
	\begin{subfigure}[c]{0.48\textwidth}
		\centering
		\includegraphics[width=\textwidth]{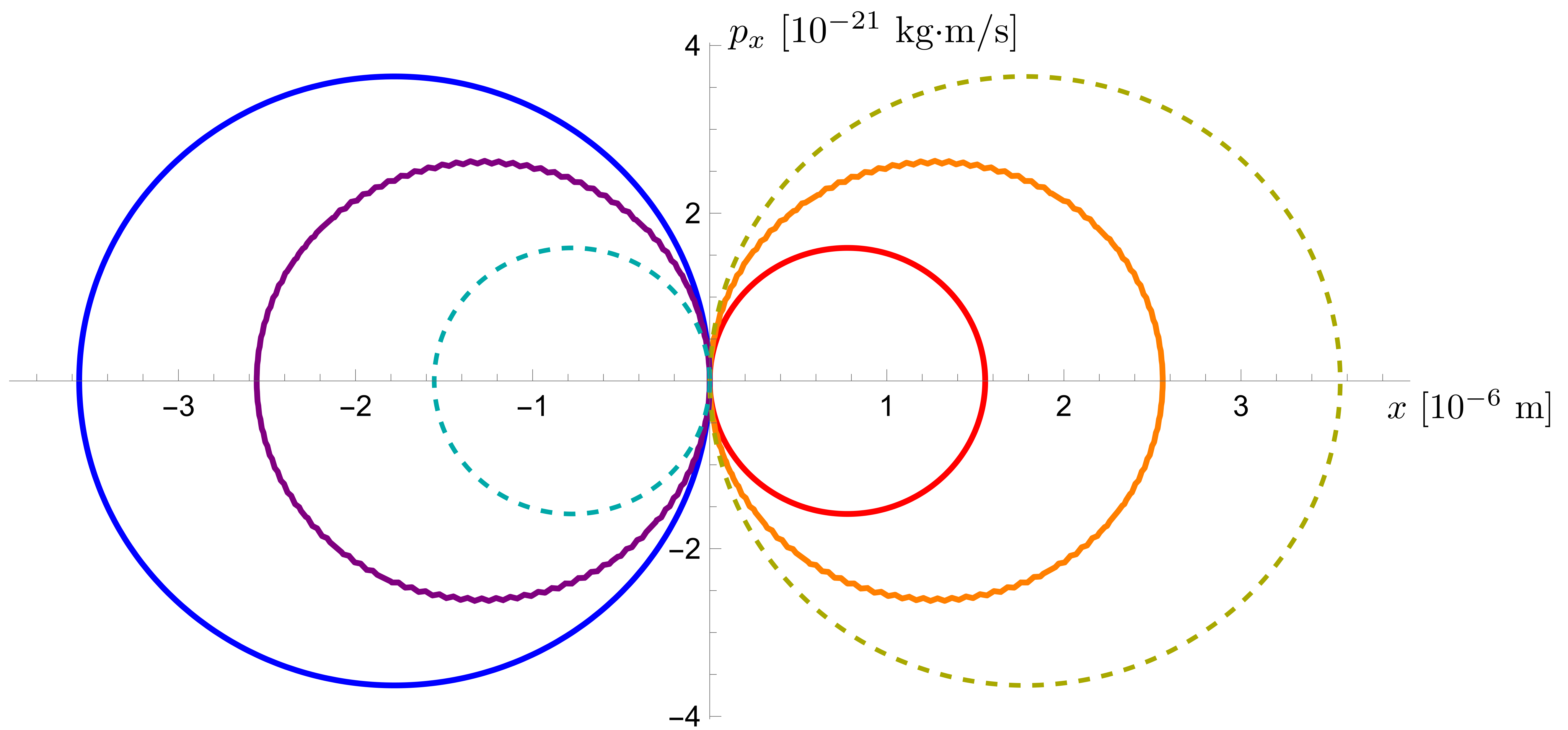}
		\subcaption{$B_0=5\times10^{-4}$ T; $\omega_{DD}=200\omega$.}
		\label{NN200}
	\end{subfigure}
	\caption{Plots (a-c): Phase space diagram of the ND superposition components, in absence of DD and for two different $\omega_{DD}$ values. In plots (b) and (c), red, green-dashed and orange curves, respectively: dynamics generated  by $H_x^{(\uparrow)}$, $H_x^{(\downarrow)}$ and $H_x^{(DD)}$ for $S_x=+1$.
Blue, azure-dashed and purple curves, respectively: dynamics generated by $H_x^{(\uparrow)}$, $H_x^{(\downarrow)}$ and $H_x^{(DD)}$ for $S_x=-1$. See Appendix for further details.}
\label{phase_DD}
\end{figure}

\textit{DD with separate control of $B_0$ and $B'$ - } An alternative configuration could involve separate control of $B'$ and $B_0$ (the latter generated by adding a Helmholtz coil or two permanent magnets to the setup), allowing to eventually flip only $B'$ together with $S_x$ while leaving $B_0$ untouched.
Given the peculiar form of the Hamiltonian in Eq. \eqref{ham}, this is equivalent to a scenario in which $B'$ and $S_x$ are left unperturbed and only $B_0$ flips, leading to the time-dependent Hamiltonian:
\begin{align}
H_x^{(DD)}(t)=&\sum_{j=0}^\infty\Bigg(\frac{p^2_x}{2m} + \frac{\chi V(B_0(-1)^j + B'x)^2}{2\mu_0} + \nonumber \\
 & +\hbar  \gamma_eS_x(B_0(-1)^j + B'x) + \hbar D S^2_x\Bigg)\cdot \nonumber\\
&\cdot \left[\Theta\left(t-\frac{2\pi j}{\omega_{DD}}\right)-\Theta\left(t-\frac{2\pi(j+1)}{\omega_{DD}}\right)\right]\;,
\label{hamDD}
\end{align}
being $\Theta(x)$ the Heaviside function.
Here, each superposition component will alternatively swap between two Hamiltonians of the form of Eq. \eqref{hamq}, $H_x^{(\uparrow)}$ and $H_x^{(\downarrow)}$ (with $H_x^{(\uparrow\downarrow)}=H_x(\pm B_0)$), both generating a (shifted) harmonic motion but with different equilibrium positions.
Under the Hamiltonian of Eq. \eqref{hamDD}, the initial state of our ND evolves almost like in Eq. \eqref{evol}:
\begin{align}
&\ket{\alpha(t)}_\pm\;\rightarrow\; \ket{\alpha(t)}_\pm^{(DD)}=\sum_{j=0}^\infty\mathcal{C}^{(j)}_\pm(t)\cdot\nonumber\\
&\cdot\left[\Theta\left(t-\frac{2\pi j}{\omega_{DD}}\right)-\Theta\left(t-\frac{2\pi(j+1)}{\omega_{DD}}\right)\right]\ket{\alpha^{(j)}_\pm(t)}\;,
\label{evolDD}
\end{align}
with the global dynamics of the ND wavefunction components expressed as a composition of the dynamics in the single time intervals (see Appendix).\\
This is highlighted by the phase space diagram in Fig. \ref{phase_DD}, obtained considering $B_0=5\times10^{-4}$ T and different $N$ values.
%
%Indeed, in case of $N=4$ (Fig. \ref{NN4}), the trajectory of both superposition components (orange and purple curves for $S_x=1$ and $S_x=-1$, respectively) initially follows the one dictated by the Hamiltonian in Eq. \eqref{hamq}, corresponding to $H_x^{(\uparrow)}$ in the DD regime (red and blue curves, respectively).
%Then, in correspondence of the first $\pi$ pulse, they start to feel a different harmonic potential, and their trajectory becomes a circle centered on the same equilibrium position as the trajectory dictated by the $H_x^{(\downarrow)}$ Hamiltonian (dashed green and azure curves, respectively).
%The subsequent $\pi$ pulse restores $H_x^{(\uparrow)}$, hence each ND component makes an arc centered on its initial equilibrium position until the following pulse, that makes both $S_x$ components follow the trajectory stemming from $H_x^{(\downarrow)}$ to return to their initial position.\\
%For increasing $\omega_{DD}$, e.g. for $N=20$ as in Fig. \ref{NN20}, the frequent flipping between the two Hamiltonians starts to highlight some ``gearwheel'' dynamics, that seems to approach the one of a harmonic oscillator with an average equilibrium position between the ones given by $H_x^{(\downarrow)}$ and $H_x^{(\uparrow)}$.
%This is confirmed by the asymptotic behavior obtained for $\omega_{DD}\gg\omega$ (i.e., for the scenario of an actual DD implementation), as shown in Fig. \ref{NN200}.\\
The figure shows how, for small $N$, the ``jumps'' between $H_x^{(\uparrow)}$ and $H_x^{(\downarrow)}$ appear in both trajectories (orange and purple curves for $S_x=+1$ and $S_x=-1$, respectively), approaching the one of a harmonic oscillator with an ``average'' equilibrium position for large $N$ (i.e., in the actual DD scenario).
Remarkably, even though the non-negligible $B_0$ considered here already unbalances the trajectories of the two ND superposition components without DD (or for $B_0$ flipping accordingly with $\omega_{DD}$), shifting their equilibrium positions (as it can be seen in Fig. \ref{NN0}), the trajectories dictated by $H_x^{(DD)}$ are symmetric with respect to the origin, as it was before for $B_0=0$ (see Fig. \ref{mm2.9}).
This means that DD makes our ND superpositions unperturbed by residual (undesired) bias fields in our setup.\\% (in this case, to characterize the ND superposition one could initialize the NV-centre $S_x$ state in one of the $\ket*{\psi_{0\pm}}=\frac{1}{\sqrt2}(\ket*{0}\pm\ket*{1})$ states instead of $\ket*{\psi_S}$).\\
\section{Table-top quantum gravity experiment}

\rev{The main advantage offered by our proposal is that our setup exploits partially confined quantum probes instead of free-falling ones, specifically two spatially-delocalized ND superpositions hosted in the same trap (see Fig. \ref{scheme}).
For this purpose, single-NV-centre NDs with near-bulk properties and high (DD-enhanced) coherence time will be fabricated \cite{And14,march2023long} and characterized by means of single-photon Hanbury-Brown \& Twiss interferometry \cite{Gat2014} and optically-detected magnetic resonance (ODMR) \cite{Gru97,Doh13}.}\\
Selected NDs will be ejected from the growth array via a blister-based laser-induced forward transfer (LIFT) technique \cite{Kom2020}, essentially pushing the ND from the donor (the array) to the acceptor (the trap).
A 2D magnetic potential will confine the NDs in $y$ and $z$, while the position of each ND in the (unconstrained) $x$ direction can be controlled, e.g., with doughnut-shaped optical tweezers, to reduce light absorption by the trapped NDs and avoid thermal damage in vacuum \cite{Zho2017}.\\ 
\rev{Given the weakness of gravity, it is crucial to avoid mutual interactions between the NDs that might compete with it, i.e. magnetic and electric dipole-dipole coupling and Casimir-Polder forces.
Their role has been already discussed in the literature, showing how adding a Faraday-Meissner shield between the NDs can in principle reduce magnetic and electric interactions \cite{Bos2017,Kam2020,sch2023,sch2024,ela2025,Ped2020} which, otherwise, would overwhelm the gravitational one by orders of magnitude.
However, introducing such a plate generates in turn new Casimir-Polder forces and other noise sources \cite{bul26}, for which accurate additional calibration is required.
Casimir-Polder interactions are the strongest at distance $d\lesssim10$ $\mu$m, with a fast decay $\propto d^{-7}$.
For our purpose, we assume as a criterion to determine the minimal ND-ND initial distance $d_\text{min}$ the one granting a minimum separation $\Delta_\text{CP}$ between the closest ND superposition components granting a Casimir-Polder potential $V_\text{CP}$ one order of magnitude below the gravitational one ($V_G$), i.e.:
\begin{equation}\label{dmin}
    d_\text{min}=\Delta x_\text{max}+\Delta_\text{CP}\;\;\;\;\;\;\Delta_\text{CP} \approx \left[\frac{\hbar cV^2}{Gm^2}\left(\frac{\varepsilon-1}{\varepsilon+2}\right)^2\right]^\frac16,
\end{equation}
determined by the nanoparticle density and relative permittivity and yielding, for NDs, $\Delta_\text{CP}\simeq 160~\mu$m. }

\begin{figure}[th]
	\centering
	\includegraphics[width=\columnwidth]{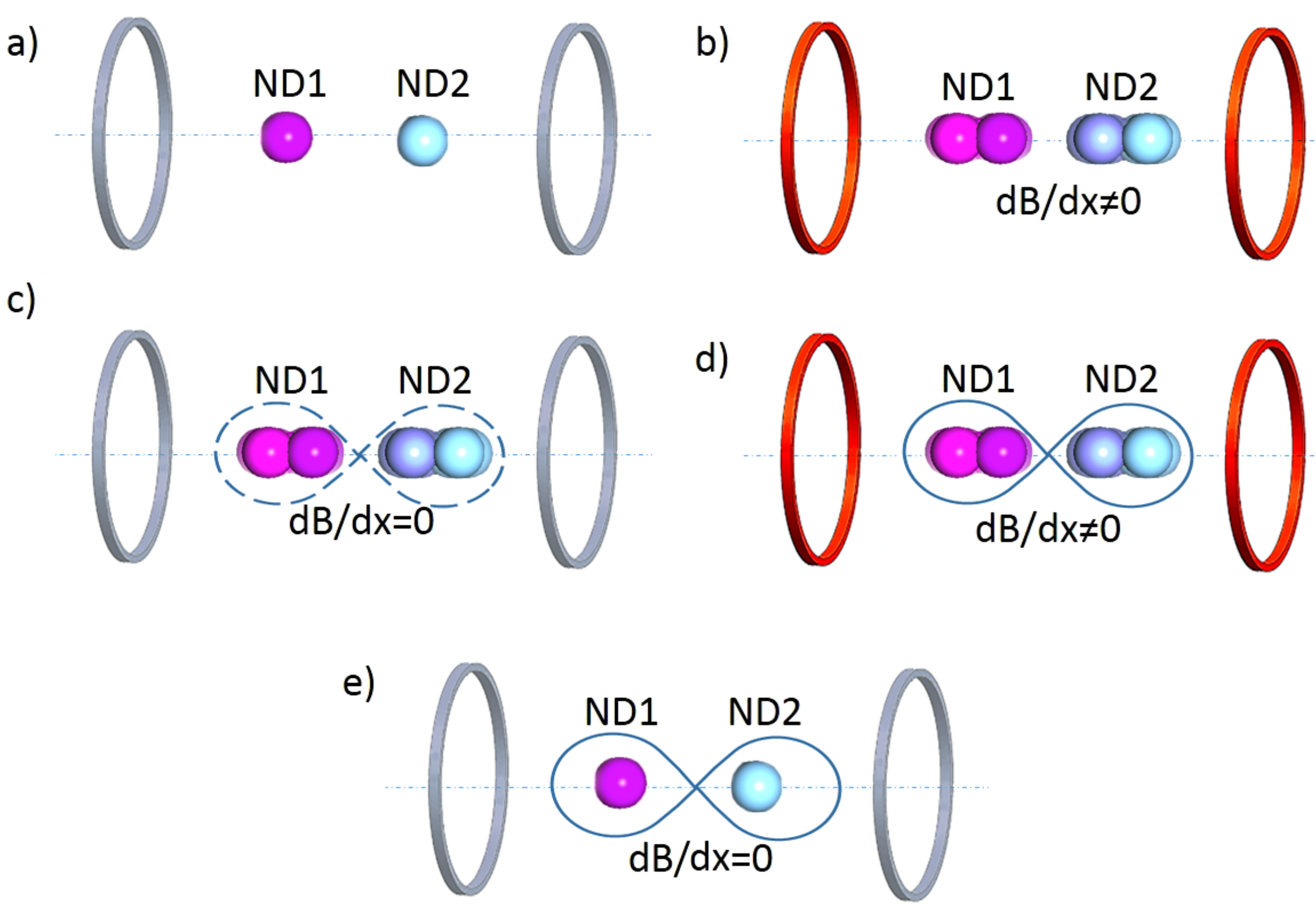}
	\caption{Step-by-step scheme of the ND-based experiment unveiling quantum features in gravity: a) two single-NV-centre NDs are confined in the same trap along $y$ and $z$, leaving $x$ unconstrained; b) a magnetic gradient $B'=dB/dx$ is generated, putting both NDs in a spatially-delocalized superposition; c) $B'$ is switched off, leaving the two superpositions freely-evolving by gravitational interaction; d) $B'$ is turned on again, to recombine both spatial superpositions; e) $B'$ is finally turned off, and a correlated measurement of the two NV-centres is performed to detect gravity-induced entanglement.}
	\label{exp}
\end{figure}

\begin{figure}[th]
	\centering
	\includegraphics[width=\columnwidth]{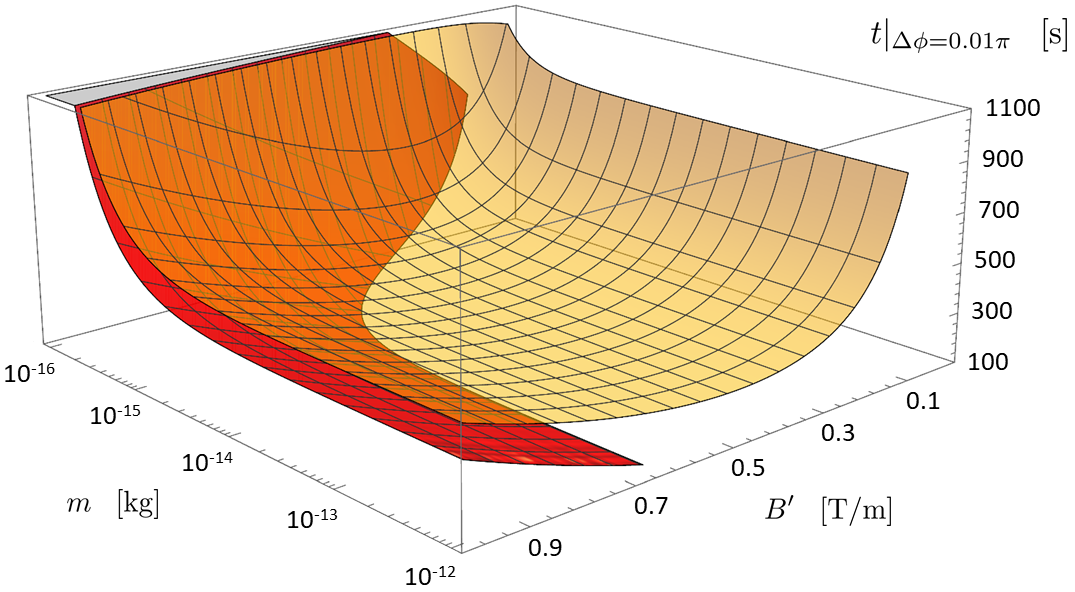}
	\caption{Yellow surface: time needed to complete the entire experimental protocol of Fig. \ref{exp} and entangle the two NDs by accumulating a gravity-induced phase $\Delta\phi=0.01\pi$ in the (c) step, as a function of the NDs mass $m$ and the magnetic field gradient $B'$, considering the NDs initially put at the minimum distance $d_\text{min}$ of Eq. \eqref{dmin}, allowing to keep the Casimir-Polder potential one order of magnitude below the gravitational one. Red surface: time needed to complete the same protocol and achieve a gravity-induced phase $\Delta\phi=0.01\pi$ in the ND bipartite state considering also the progressive phase accumulation achieved in the (b) and (d) steps of the protocol, as a function of the NDs mass $m$ and the magnetic field gradient $B'$ and considering NDs initially put at the same minimum distance $d_\text{min}$.}
	\label{t_deltaphi}
\end{figure}

Our configuration relies on a time-varying magnetic field having a spatially-uniform gradient generated by anti-Helmholtz coils \cite{Jan2020}, with a single mW antenna for implementing DD, instead of the meters-long micro-fabricated permanent magnets \cite{Woo2022a} hosting $\gtrsim10^4$ DD antennas needed in a free-fall experiment.\\
%Instead of meters long permanent magnets \cite{Woo2022a}, our configuration relies on a time-varying magnetic field having a spatially-uniform gradient generated by anti-Helmholtz coils \cite{Jan2020}.
%Furthermore, we can implement DD with a single microwave antenna instead of the $\gtrsim10^4$ antennas needed in a free-fall experiment, requiring micro-structured fabrications for the of meters long magnets with micro-structures of the order of the $\mu$m.\\
%In refs. \cite{Bos2017,Mar2017} it was shown that, despite the weakness of gravity, eventual quantum features in the gravitational interaction between two ND spatial superpositions in adjacent matter-wave interferometers can entangle the NDs (impossible in the LOCC framework \cite{Mar2020}).
Afterwards, the quantum gravity experiment will be run as depicted in Fig. \ref{exp}.
In Fig. \ref{exp}a, we assume the two NDs to be in the same magnetic trap, separated by a distance $d$ along the $x$ axis, at low internal temperature (4 K) and pressure \rev{(both required for suppressing decoherence, as discussed in Sec. IIA).}
The single NV-centre of each ND is prepared in the state $\ket{\psi_S}$.
%Being crucial to avoid stronger-than-gravity interactions like the static \cite{Bos2017} and dynamical \cite{Bar2016} Casimir-Polder one \cite{Bos2017,Bar2016,Kam2020,Ami2023} (bounding the minimum ND-ND distance to $\sim200$ $\mu$m to keep this effect one order of magnitude below the gravitational interaction \cite{Bos2017}), we assume the two NDs to be in the same diamagnetic trap at a distance $d\sim$ 300-400 $\mu$m along $x$, at low internal temperature (4 K) and pressure (both required for suppressing decoherence), see Fig. \ref{exp}a.
Then, we achieve the maximally-separated ND spatial superposition by applying $B'$ (Fig. \ref{exp}b).
Afterwards, $B'$ is turned off in correspondence of one of the maxima of the oscillatory motion (so that the speed along $x$ of the superposition components is 0), and only mutual gravitational interaction remains  (Fig. \ref{exp}c).
The duration of this step should be adequately chosen to achieve significant entanglement while keeping a high coherence for each superposition.
Subsequently, both ND superpositions are recombined along $x$ by restoring $B'$ (Fig. \ref{exp}d).
Finally, after turning off $B'$ a single-shot measurement of the global spin state of two NDs NV-centres is performed (Fig. \ref{exp}e).\\
By repeating this procedure, we can evaluate (quantum) correlations revealing gravitationally-induced entanglement between the NDs.
Indeed, as per Ref. \cite{Bos2017}, while initially the NV centres of the two NDs form the separable bipartite state $\ket{\psi_i}=\ket{\psi_S}\otimes\ket{\psi_S}$, the mutual gravitational interaction among the different components of the spatially-delocalized ND superpositions will generate the final state:
\begin{equation}\label{ent_ND}
  \ket{\psi_f}=\frac12(\ket{-1,-1}+\ket{1,1}+e^{i\phi_-}\ket{1,-1}+e^{-i\phi_+}\ket{-1,1})\;,
\end{equation}
being $\phi_\pm=\pm\frac{G\,m^2}{\hbar}\left(\frac{1}{d}-\frac{1}{(d\pm\Delta x)}\right)t$ the gravity-induced phases during the interaction time $t$.
%Such a state is still separable for an overall phase mismatch $\Delta\phi=\phi_--\phi_+=0$, but results entangled for, e.g., $\Delta\phi>0$.\\
Such a state results entangled if the overall phase mismatch $\Delta\phi=\phi_--\phi_+=\frac{G\,m^2}{\hbar}\left(\frac{1}{(d-\Delta x)}+\frac{1}{(d+\Delta x)}-\frac2d\right)>0$, while it remains separable for $\Delta\phi=0$, i.e. for $\Delta x=0$.\\
As stated previously, NDs with $\sim250$ nm of diameter are probably not suitable for revealing gravity with this method, since their mass would be too small.
As a consequence, exploiting larger NDs implies that, to achieve a comparable separation between the two spin components, one should lower the magnetic field gradient $B'$, thus increasing the duration of the harmonic oscillation needed to separate and then recombine the two components.
To obtain the optimal trade-off between the ND mass and the magnetic field gradient, we performed a sensitivity analysis on our interferometer, obtaining the minimum time $t|_{\Delta\phi}$ (as a function of $m$ and $B'$) for running the whole protocol and achieve a gravity-induced phase $\Delta\phi=0.01\pi$ entangling the output state components, as suggested in Ref. \cite{Bos2017}.
Our calculations showed that, as expected, the minimum time is obtained when the initial distance $d$ between the two NDs is the minimum achievable, i.e. $d=d_\text{min}$.
The yellow surface reported in Fig. \ref{t_deltaphi} shows the time needed to complete the entire protocol of Fig. \ref{exp} and accumulate an overall gravity-induced phase $\Delta\phi=0.01\pi$ in the (c) step of protocol, as a function of the NDs mass $m$ and the magnetic field gradient $B'$ - these are the only free parameters of the system, once we fix the diamond density to be 3550 kg/m$^3$.
%
%\begin{figure}[th]
%	\centering
%	\includegraphics[width=\columnwidth]{t_deltaphi_0.01pi.pdf}
%	\caption{\rev{Time needed to complete the entire experimental protocol of Fig. \ref{exp} and entangle the two NV centres with a gravity-induced phase $\Delta\phi=0.01\pi$, as a function of the NDs mass $m$ and the magnetic field gradient $B'$, considering the NDs initially put at the minimum distance $d_\text{min}$ of Eq. \eqref{dmin}, allowing to keep the Casimir-Polder potential one order of magnitude below the gravitational one.}}
%	\label{t_deltaphi}
%\end{figure}
The plot highlights that the minimum protocol duration becomes (almost) mass-independent for $m\gtrsim10^{-14}$ kg; this is due to the fact that, when the NDs mass grows, the advantage given by the stronger gravitational interaction is counterbalanced by the slower ND superposition states separation/recombination stages.
The minimum time needed to complete this experimental protocol, corresponding to $m\simeq10^{-12}$ kg and $B'=0.475$ T/m, is $t_{min}|_{\Delta\phi=0.01\pi}\simeq283$ s.\\
This time interval can be significantly diminished if we consider also the gravity-induced phase accumulated during the NDs spatial superposition states separation and recombination steps.
In such a scenario, corresponding to the red surface of Fig. \ref{t_deltaphi}, we have a smaller region of available $m$ and $B'$ combinations, since we have to satisfy the boundary condition of the minimum protocol duration corresponding to an entire period of the NDs oscillatory motion.
%
%\begin{figure}[th]
%	\centering
%	\includegraphics[width=\columnwidth]{t_deltaphi_0.01pi_tot.pdf}
%	\caption{\rev{Time needed to complete the entire experimental protocol of Fig. \ref{exp} and entangle the two NV centres with a gravity-induced phase $\Delta\phi=0.01\pi$, as a function of the NDs mass $m$ and the magnetic field gradient $B'$, considering also the progressive phase accumulation achieved in the (b) and (d) steps of the protocol. Again, the initial distance between the NDs is $d_\text{min}$ i, i.e. the minimum one granting a Casimir-Polder potential one order of magnitude below the gravitational one.}}
%	\label{t_deltaphi2}
%\end{figure}
%
By comparing these two surfaces, one can immediately guess that, if we consider also the phase accumulated during the NDs spatial delocalization/recombination steps, the trade-off between stronger gravitational interaction (granted by larger NDs mass $m$ and smaller magnetic field gradient $B'$) and shorter duration of the interferometric protocol (achieved for larger $B'$) allows reaching, for $B'=0.663$ T/m and $m\simeq5.6\times10^{-14}$ kg, $t_{min}|_{\Delta\phi=0.01\pi}\simeq135$ s, less than half of the one expected for the scenario in which only the phase accumulated in step (c) of the protocol is considered.
%\rev{Proposals have been made for further shortening this time span, e.g. by decreasing $\Delta_\text{CP}$ with the insertion of a conducting plate for screening the Casimir-Polder effect \cite{sch2023}, although this might introduce further noise sources \cite{bul26}.}\\
\rev{The aforementioned possibility of inserting a conducting plate for screening the Casimir-Polder effect \cite{sch2023} might reduce $\Delta_\text{CP}$ and, as a consequence, $t_{min}$, although at the cost of introducing further noise sources \cite{bul26}, as previously mentioned.}\\
%Our calculations, run by considering a separation $d=d_\text{min}$ between our NDs, showed that such a $\Delta\phi$ value could be obtained for $m\sim10^{-14}-10^{-12}$ kg with $B'\simeq0.2$ T/m, fixing the overall protocol duration at $t|_{\Delta\phi}\sim500$ s.
%Such a value can be decreased to $t|_{\Delta\phi}\sim200$ s if we consider also the phase accumulation during the separation and recombination steps, allowing us to exploit $B'\sim0.3$ T/m.\\
The problem of the NV centre coherence time (that, anyway, holds also for schemes exploiting free-falling/travelling objects, such as those of Refs. \cite{Mar2017,Bos2017}) remains one of the main challenges for the realization of this experiment.
Nevertheless, several ideas for improving this aspect already exist, both for NV centres \cite{Bar2013,Hens2023,Lei2025,Tan2025,Lo2025} or group IV color centres \cite{Gu2025} suited for this scheme \cite{Ros2024,Brandenburg2018}.
\rev{While bulk diamond samples have recently demonstrated coherence times approaching the required timescales \cite{yamamoto2026ten}, as previously stated translating this performance to NDs remains challenging because of the dense bath of surface spins, although on-purpose fabrication techniques can overcome this limitation \cite{And14}.
Conversely, center-of-mass decoherence imposes arguably stricter operational limits.
As discussed in Section IIA, mitigating collisional and thermal decoherence requires operating in XHV regimes (already achieved at collider facilities \cite{PhysRevLett.82.3198,Benvenuti1993}), alongside the development of highly pure, ultra-low-absorption ND samples \cite{Frangeskou2018} to suppress blackbody photon scattering.\\
To eventually reduce the overall execution time of the experiment, protocols for fast \cite{Mar2022, Zhou2022} and mass-independent \cite{Mar2022,Zhou2023} expansion of the motional wavepacket have been proposed, although the required fast-switching control elements would inevitably introduce additional electronic noise. Furthermore, pushing the wavepacket into the non-linear regions of the trapping potential during rapid, wide expansions might induce motional non-Gaussianity or ND loss \cite{Wu2024}.\\
For the sake of completeness, we performed a sensitivity and robustness analysis of the proposed ND interferometer for the gravity-induced entanglement verification scenario (detailed results in Appendix B).
As a first step, we studied the case in which the ND is not put exactly in the origin of our reference frame ($x=y=z=0$). %, but in a point belonging to the surface of a sphere (diameter: 1 $\mu$m) centered in the origin.
Our numerical simulations show that, for a shift in the $x$ axis, the oscillatory motion of the ND superposition components might change, but not their maximal separation.
If the shift occurs in the $z$ or $y$ axis, instead, an additional oscillatory motion in these directions is induced, but identical for both the ND superposition components, leaving the dynamics along the $x$ axis unchanged.\\
Regarding the DD application, considering the fact that seconds long DD sequences might suffer from time delays (introducing
additional dephasing), we studied the effect of a possible phase shift $\delta$ between the flipping of the magnetic field gradient along $x$ and the microwave $\pi$ pulse inverting $S_x$.
The results of our analysis, detailed in the Appendix, show that for small phase shifts ($\delta<\frac{\pi}{15}$) the trajectories of the ND superposition components are almost unaffected, with a small but still noticeable variation of the oscillatory behavior for $\delta\gtrsim\frac{\pi}{5}$.

Altogether, our scheme provides a promising route toward next-gen table-top tests of gravity-induced entanglement between superposed massive objects, offering several practical advantages over free-flying configurations: e.g., switching from a meters long micro-featured structure to a more compact table-top setup, with a much easier low-temperature and vacuum management (not to mention the vastly more efficient and controllable DD implementation).
Furthermore, after the final measurement on their NV centres, here we do not lose the NDs, allowing us to recycle them without the need to find and characterize new samples for each run (a time-consuming task that would heavily slow down the experiment), intrinsically avoiding ``noise'' arising from fabrication discrepancies among the several ND samples required by previous schemes.\\
In the single-ND configuration, the apparatus serves as a versatile platform for matter-wave experiments, where the delocalization of large masses can be directly probed using a Ramsey interferometry protocol.
We here emphasize that this platform approaches technological readiness, assuming a streamlined and reliable production of NDs hosting single NV centers.
Owing to its intrinsic layout, the setup is naturally insensitive to homogeneous bias magnetic fields and remains stable against alignment errors or arising from the use of different samples.
%Additional control capabilities—such as spin initialization and manipulation, feedback cooling of the center-of-mass motion, and spin and motional DD sequences to prolong coherence can be straightforwardly implemented in XHV environments.
When two NDs are employed, the coil arrangement supports inertial free-fall–like motion along the $x$ axis, in principle enabling the detection
of gravity-induced phases as well as generic mutually-induced phases, as it permits:
(i) generating large superpositions of massive objects, 
(ii) keeping the samples at a fixed separation for an adjustable time interval, and 
(iii) accessible readout via joint spin measurements.\\
Finally, the versatility of the proposed setup allows straightforward implementation of the additional elements discussed above (transient magnetic fields, shielding plates), which may help to overcome current
limitations.}

\section{Concluding remarks}

The potential observation of gravity-induced entanglement (impossible to stir by LOCC) between two spatially-superposed nano-masses \cite{M17,Mar2017,Bos2017} would represent a breakthrough in physics, since it would become a first hint of the quantum nature of gravitational interactions \cite{MVr} as well as the cornerstone of experimental quantum gravity \cite{Add2022} (as mentioned, such experiments would also allow testing macro-objectivation \cite{Gen2010} and, in particular, dynamical collapse models \cite{Bas2023}).\\
\rev{A key innovation of our approach lies in the possibility of applying both spin- and motional- DD \cite{Ped2020} to extend the coherence time, shielding the system from homogeneous bias fields.
Our analysis reveals that, in the single-ND configuration, it is possible to generate and observe macroscopic superposition (with spatial separation around one order of magnitude above the ND diameter) for samples with $m\sim10^{-17}$ kg using nearly-available technology \cite{Pedalino2026,Bild2023,Bateman2014}.
The remaining challenge for the experiment realization is the reliable fabrication of highly pure NDs hosting single NV centers with coherence times under dynamical decoupling $T_2^*\geqslant 100$ ms, which is already the case for bulk diamonds. 
As double-ND experiments revealing gravity-induced entanglement typically require larger masses scaling up the strength of the gravitational potential, we discuss a realistic implementation of the BMV protocol \cite{Bos2017,Mar2017} contextual to our setup employing NDs with $m\gtrsim10^{-14}$ kg, proving it as a robust testbed for the detection of gravity-induced entanglement.\\
%We illustrate the ND superposition components behavior in different scenarios, showing how the implementation of a DD mechanism, besides extending the NV-centre coherence time, could make our quantum superposition robust against potential biases induced by stationary fields.
%By adding a second spatially-delocalized ND to the magnetic trap and regulating the distance between the two NDs, our table-top system could enable experiments aiming at revealing quantum features in gravity by detecting gravity-induced entanglement (impossible to stir by LOCC) between the two NDs.
In contrast to free-falling proposals \cite{Bos2017,Mar2017,MVr}, by maintaining the NDs in a stable levitated potential our scheme allows re-initializing and re-using them for several consecutive measurement runs (removing the need to find and trap new NDs run by run), dramatically increasing the data collection rate and bringing the accumulation of the necessary
statistics significantly closer to experimental feasibility.\\}
Finally, being sensitive to extremely weak external fields, this kind of interferometer would also qualify as an innovative and very powerful quantum sensor.

%Our analysis demonstrates the feasibility of this set-up, paving the way to a realistic realisation of this experiment.

%\section*{Acknowledgments}
\begin{acknowledgments}
This work was financially supported by the project QuteNoise (call ``Trapezio'' of Fondazione San Paolo), by the Qu-Test project, which has received funding from the European Union's Horizon Europe Research and Innovation Programme under grant agreement No. 101113901, and by the project 23NRM04 NoQTeS.
The project 23NRM04 NoQTeS has received funding from the European Partnership on Metrology, co-financed from the European Union's Horizon Europe Research and Innovation Programme and by the Participating States.
This work was also funded by the MUR project AQuTE, grant No. 2022RATBS4 (call PRIN 2022), and by the INFN project CSN5 QUISS.
MB acknowledges support by the Horizon Europe EIC-Pathfinder project QuCoM (grant agreement No. 101046973).
We thank Chiara Marletto and Vlatko Vedral for fruitful discussions.\\
\end{acknowledgments}

\appendix

\section{Dynamics of a single-NV-centre ND in the table-top interferometer}
\label{Appendix: Dynamics}

\begin{figure}[th]
	\centering
	\includegraphics[width=0.5\columnwidth]{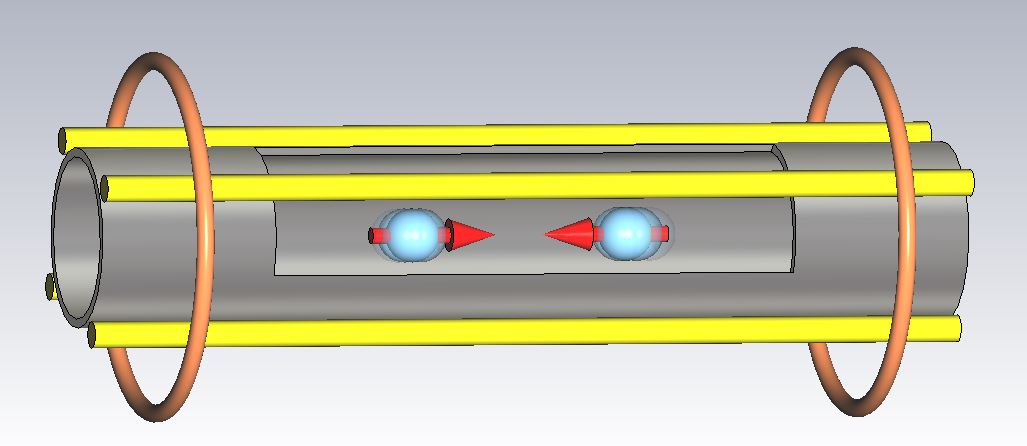}
	\caption{The physical system under study is a single-NV-centre nanodiamond (trapped in the $y$ and $z$ directions by a gravito-magnetic potential) in a spatially-delocalized superposition, generated by putting the NV center spin component $S_x$ in the initial state $\ket*{\psi_S}=\frac{1}{\sqrt2}(\ket*{-1}+\ket*{1})$ before applying a magnetic field gradient $B'$ along the $x$ direction, thus spatially separating the two components of $\ket*{\psi_S}$. Depending on its $S_x$ orientation (indicated by the red arrow), each component will then oscillate in the $x$ direction around a different equilibrium position.}
	\label{scheme1}
\end{figure}
Case-study (Fig. \ref{scheme1}): stationary single-NV-centre ND free to move along the $x$ direction while being constrained in the $y$ and $z$ ones, with the NV-centre spin axis aligned along the unconstrained direction, confined by exploiting a magnetic trap \cite{Pri1983,Tre2004b,For2007,Bau2020} with a 2D confining magnetic potential in $y$ and $z$.
The $S_x$ component of the ND NV-center is initialized in the state $\ket*{\psi_S}=\frac{1}{\sqrt2}(\ket*{-1}+\ket*{+1})$ by a sequence of optical pumping followed by a $\pi$-pulse. Due to the low-frequency regime of magnetic trapping, additional spin-based feedback cooling \cite{Delord2020} must be applied to reduce the center of mass phononic occupation. Under these premises, the initial state reads can be assumed to be: 
\begin{equation}
\ket{\Psi_I}=\ket{\psi_S}\otimes\ket{0},
\label{psiI}
\end{equation}
where $\ket{0}$ is the spatial component of the ND wave function.\\
To create a spatially-delocalized ND superposition we generate a magnetic field gradient $B'=\frac{dB}{dx}$, giving rise (along the $x$ direction) to the Hamiltonian of Eq. \eqref{ham}.
%
%\begin{equation}
%H_x=\frac{p^2_x}{2m} + \frac{\chi V(B_0 + B'x)^2}{2\mu_0} + \hbar\gamma_eS_x(B_0 + B'x) + \hbar D S^2_x\;,
%\label{ham}
%\end{equation}
%
%corresponding Eq. (1) of the main paper, hosting (respectively): $m$, the ND mass; $p_x$, its momentum along $x$; $V$, its volume; $\chi$, its magnetic susceptibility (assuming for the ND a diamagnetic behavior); $\mu_0$, the vacuum magnetic permeability; $\gamma_e$, the gyromagnetic ratio of the electron; $B_0$, the bias magnetic field; $D$, the zero-field splitting of the NV centre.
%Classically, the dynamics generated by this Hamiltonian corresponds, for two separated particles of opposite spin, to a harmonic oscillation with the same frequency but different equilibrium positions:
%%
%\begin{equation}
%x^{(\pm)}(t)= x_0^{(\pm)}(1-\cos(\omega t)),
%\label{motion}
%\end{equation}
%%
%being $x_0^{(\pm)}=-\frac{\chi V B_0\pm\hbar\gamma_e\mu_0}{\chi V B'}$ the equilibrium position of the particle corresponding to the $S_x=\pm1$ spin component, and $\omega=B'\sqrt{\frac{\chi V}{\mu_0m}}$ the oscillation frequency.\\
\rev{This Hamiltonian can be recasted in a quantum harmonic oscillator form by expanding it as
%
%\begin{equation}
%    H = \frac{\hp^2}{2m} + \frac{1}{2}m\omega^2\hx^2 + \hbar\left( \frac{\chi V }{\hbar\mu_0}B_0B' + \gamma_e B' \hS \right) x+ \hbar\gamma_eB_0 \hS +\hbar D \hS^2\;,
%\end{equation}
%
\begin{align}
    \hat{H}_x =& \frac{\hp_x^2}{2m} + \frac{1}{2}m\omega^2\hx^2 +\nonumber\\
    &+ \hbar\left( \frac{\chi V }{\hbar\mu_0}B_0B'+\frac{mg\sin\theta_g}{\hbar} + \gamma_e B' \hS \right) \hat{x}+ \nonumber\\  
    &+ \hbar\gamma_eB_0 \hS +\hbar D \hS^2
\end{align}
where the harmonic frequency $\omega= B'\sqrt{\frac{\chi V}{\mu_0m}}$ depends only on the magnetic field gradient, and the properties of the ND.
We can thus define normal modes of oscillation of the spin center of mass (CoM) $\hx= x_\text{zpf}(\ha+\had)$ and $\hp_x= ip_\text{zpf}(\had-\ha)$, where $x_\text{zpf}=(\hbar/2 m \omega)^{1/2}$ and $p_\text{zpf}=\left({\hbar m \omega /2}\right)^{1/2}$ are the zero-point fluctuations standard deviations, and $\ha, \had$ are the annihilation and creation bosonic operators for the quantized phononic oscillation modes of the center of mass, satisfying $[\ha,\had]=1$.
In these terms, the Hamiltonian of the systems reads:
\begin{equation}
    \begin{aligned}
    \label{equ:QHOHamiltonian}
        H = \hbar\omega \had\ha + \hbar\left(\lambda_0+\lambda \hS\right)(\ha+\had) + \hbar\gamma_e B_0\hS+\hbar D \hS^2\;,
    \end{aligned}
\end{equation}
with homogeneous and spin-dependent coupling rates $\lambda_0 = ( \chi V B_0B' / \mu_0 +mg\sin\theta_g)x_\text{zpf}/\hbar $ and $\lambda=\gamma_e B' x_\text{zpf}$ respectively.
The first term accounts for the kinetic and harmonic energy of the system, the second for the magneto-mechanical coupling, while the third is a Zeeman splitting due to the bias field $B_0$.
The last term account for zero-field splitting of the NV center.}

The Hamiltonian in Eq.~\eqref{equ:QHOHamiltonian} can be decomposed in a sum over the block-diagonal Hilbert space of the NV, spanned by the spin eigenstates $\lbrace \ket{0},\ket{+1},\ket{-1}\rbrace$.
This can be accomplished considering the spin-eigenspace spectral decomposition
\begin{equation}
\label{equ:BlockDiagHamiltonian}
    \hat{H}=\sum_{j=0,\pm} \hat{H}_j\otimes\ket{j}\bra{j}
\end{equation}
where
\begin{equation}
    \label{equ:SingleSpinHamiltonian}
    \hat{H}_j = \hbar\omega \had\ha + \hbar\lambda_j(\ha+\had) + \hbar\gamma_e B_0s_j+\hbar D s_j^2
\end{equation}
indicate the dynamics generated by the single spin eigenvalue $s_j=\{0,\pm1\}$ for $j=\{0,\pm1\}$, and with respective coupling strength $\lambda_j = \lambda_0+\lambda s_j$.

To better understand the dynamics, we first note that the different diagonal blocks commute with each other, thus generating independent processes.
This implies that each $H_j$ fully characterizes the behavior of the corresponding process.
In addition, we can simplify each $H_j$ by means of the displacement operator $\hat{D}(\alpha)=\text{exp}\left( \alpha \had - \alpha^* \ha\right)$ by noticing that:
%
%\begin{equation}
%\label{equ:Hjdisplaced}
%    H_j = \hbar\omega \left( \left(\had+\frac{\lambda_j}{\omega}\right)\left(\ha+\frac{\lambda_j}{\omega}\right) -\frac{\lambda_j^2}{\omega^2} +\frac{\gamma_e B_0 s_j}{\omega}+\frac{D s_j^2}{\omega}\right)= D^\dagger(\lambda_j/\omega)\left[\hbar\omega\left(\had\ha -\frac{\lambda_j^2}{\omega^2} +\frac{\gamma_e B_0 s_j}{\omega}+\frac{D s_j^2}{\omega}\right) \right]D(\lambda_j/\omega)\;.
%\end{equation}
%
\begin{align}
\label{equ:Hjdisplaced}
    \hat{H}_j &= \hbar\omega \left( \left(\had+\frac{\lambda_j}{\omega}\right)\left(\ha+\frac{\lambda_j}{\omega}\right) -\frac{\lambda_j^2}{\omega^2} %+\right.\nonumber\\
    %&\left.
    +\frac{\gamma_e B_0 s_j}{\omega}+\frac{D s_j^2}{\omega}\right)\nonumber\\
     &=\hbar\omega \hat{D}^\dagger\left(\frac{\lambda_j}{\omega}\right)\left[\had\ha -\frac{\lambda_j^2}{\omega^2} %+\right.\right.\nonumber\\
     %&\left.\left.
     +\frac{\gamma_e B_0 s_j}{\omega}+\frac{D s_j^2}{\omega}\right]\hat{D}\left(\frac{\lambda_j}{\omega}\right)\;.
\end{align}
This allows to directly obtain the solution to the stationary Schr\"odinger equation $H_j\ket{\psi_j}=E_j\ket{\psi_j}$,
%\begin{equation}
%    H_j\ket{\psi_j}=E_j\ket{\psi_j}\;,
%\end{equation}
which, by plugging in $H_j$ in the form of Eq. \eqref{equ:Hjdisplaced}, can be rewritten as:
\begin{align}
    \hat{H}_j\ket{\psi_j}&=\hbar\omega\left(\had\ha - \frac{\lambda_j^2}{\omega^2} +\frac{\gamma_e B_0 s_j}{\Omega}+\frac{D s_j^2}{\omega}\right) \hat{D}\left(\frac{\lambda_j}{\omega}\right)\ket{\psi_j} \nonumber\\
    &= E_j \hat{D}\left(\frac{\lambda_j}{\omega}\right)\ket{\psi_j}\;.
\end{align}
Knowing that the eigenvectors of an harmonic oscillator are the Fock states $\ket{n}$, it is straightforward to see that the eigenvectors and eigenvalues solving these equations are:
\begin{align}
        \ket{\psi_j} &= \hat{D}\left(-\frac{\lambda_j}{\omega}\right)\ket{n}\;,\\
        E_j &= \hbar\omega\left( n- \frac{\lambda_j^2}{\omega^2} +\frac{\gamma_e B_0 s_j}{\omega}+\frac{D s_j^2}{\omega}\right)\;,
\end{align}
i.e., the eigenstates are displaced Fock states and the eigenvalues are the ones of a harmonic oscillator, but shifted by the spin-mechanical coupling, the Zeeman coupling with $B_0$ and the zero-field splitting of the NV.
\rev{We underline that for $\theta_g=0 \rightarrow \Delta E = E_+ - E_- =0$, indicating how in the stationary dynamics the degeneracy between $\ket{\pm1}$ is not lifted by the magnetic field, on the contrary of what usually happens with bulk NV.} 

\begin{figure}[t]
	\begin{subfigure}[c]{0.48\textwidth}
		\centering
		\includegraphics[width=\textwidth]{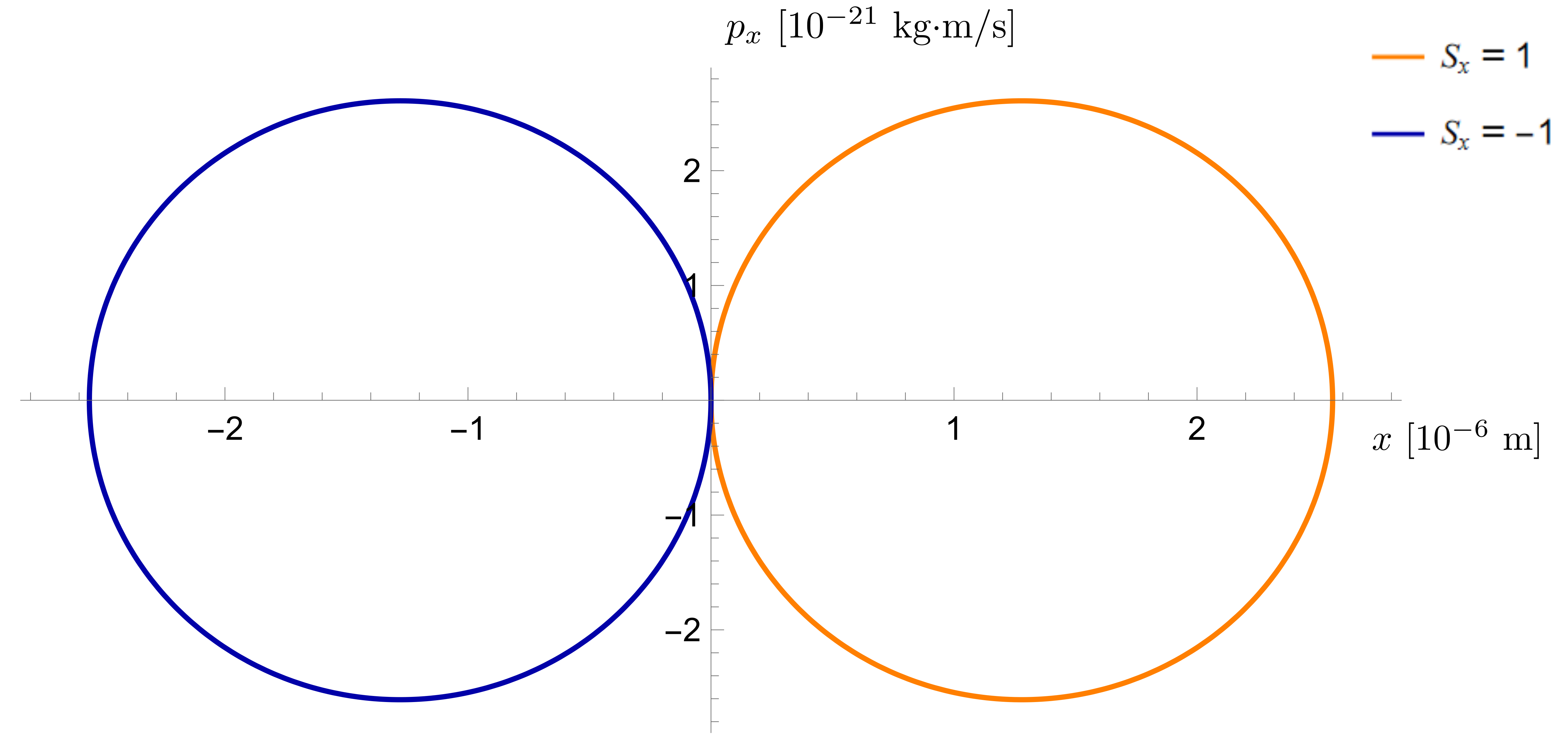}
		\subcaption{$B_0=0.$}
		\label{low}
	\end{subfigure}
	\hfill
	%\begin{subfigure}[c]{0.48\textwidth}
%		\centering
%		\includegraphics[width=\textwidth]{phase_noDD_B10-4a.png}
%		\subcaption{$B_0=10^{-4}$ T.}
%		\label{medium}
%	\end{subfigure}
%	\hfill
	\begin{subfigure}[c]{0.37\textwidth}
		\centering
		\includegraphics[width=\textwidth]{phase_noDD_B10-3.3new1.png}
		\subcaption{ $B_0=5\times10^{-4}$ T.}
		\label{strong}
	\end{subfigure}
	%\caption{Plots (a-c): Phase space diagram %($X$- and $Y$-axis, respectively: position and momentum along the $x$ direction, expressed in SI units)
%of the two ND superposition components, for different values of static magnetic field $B_0$.}
\caption{Phase space diagram ($X$- and $Y$-axis, respectively: position and momentum along $x$)
of the ND superposition components, for different values of the bias magnetic field $B_0$.}
\label{phase_noDD}
\end{figure}

The time-evolution propagator for each superposition component $\hat{U}_j(t)=\exp\left(-i\hat{H}_jt/\hbar\right)$ can be expressed, by using the decomposition of $H_j$ In \eqref{equ:Hjdisplaced}, as
\begin{align}
        \label{equ:TimePropagator}
        \hat{U}_j(t) &= \exp\left\lbrace -\frac{it}{\hbar} D^\dagger\left(\frac{\lambda_j}{\omega}\right)[\hbar\omega(\had\ha+...)]D\left(\frac{\lambda_j}{\omega}\right)\right\rbrace \nonumber\\
        &= D^\dagger\left(\frac{\lambda_j}{\omega}\right) e^{-i\phi_j t}e^{-i\omega \had\ha t}D\left(\frac{\lambda_j}{\omega}\right)\;,
\end{align}
where we have introduced the stationary phase $\phi_j = -\lambda_j^2/\omega + \gamma_eB_0s_j+Ds_j^2$ and used the property for unitary operators $\text{exp}\left(\hat{U}^\dagger\hat{O}\hat{U}\right) = \hat{U}^\dagger e^{\hat{O}}\hat{U}$.
Given the commutativity of elements of the Hamiltonian decomposition in Eq. \eqref{equ:BlockDiagHamiltonian}, the overall ND time-evolution operator can be factorized in the form:
\begin{align}
\label{U}
        \hat{U}(t) &= \text{exp}\left\lbrace -\frac{it}{\hbar}\sum_{j=0,\pm}H_j\otimes\ket{s_j}\bra{s_j} \right\rbrace \nonumber\\
        &= \prod_{j=0,\pm}\text{exp}\left\lbrace -\frac{it}{\hbar}H_j\otimes\ket{s_j}\bra{s_j} \right\rbrace \\
        &= \prod_{j=0,\pm}\text{exp}\left\lbrace -\frac{it}{\hbar}H_j \right\rbrace \otimes\ket{s_j}\bra{s_j}
\end{align}
using the idem-potency of the spin eigenstates $(\ket{s_j}\bra{s_j})^2=\ket{s_j}\bra{s_j}$
%One can then notice that the single elements of the production can be rewritten as
%\begin{equation}
%    \text{exp}\left\lbrace -\frac{it}{\hbar}H_j\otimes\ket{j}\bra{j} \right\rbrace = \mathbb{1}\otimes\mathbb{1} + \left(\mathcal{U}_j(t) -\mathbb{1}\right)\otimes\ket{j}\bra{j}
%\end{equation},
By applying it to the initial state in Eq. \eqref{psiI}, one directly obtains Eqs. \eqref{evol}-\eqref{theta_pm} from the main text.\\
In absence of a bias magnetic field ($B_0=0$), the $\ket{\alpha_\pm(t)}$ elements in Eq. \eqref{evol} describe an oscillatory dynamics with equilibrium positions symmetric with respect to the ND rest position (i.e. the ``0'' of the $x$ axis).
Introducing a bias field $B_0$ shifts concurrently equilibrium position of both $\ket{\alpha(t)}_\pm$ along $x$ in the same direction, as it can be appreciated by looking at Fig. 2 in the main paper.\\
The ND dynamics in the phase space (Fig. \ref{phase_noDD}) shows that, although for $B_0=0$ the two ND superposition components follow symmetric trajectories, for $B_0\neq0$ the equilibrium positions of both oscillatory motions are shifted of the same amount, leaving the maximum separation between them unchanged.
Specifically, Fig. \ref{low} shows the behavior of the $S_x=\pm1$ components for $B_0=0$, when they follow symmetric trajectories.
Fig. \ref{strong}, instead, highlights how, for a $B_0=5\times10^{-4}$ T, the shift in the equilibrium positions of the two oscillatory motions unbalances their dynamics.

\subsection{Checking the coherence of the ND spatially-delocalized superposition}

Straightforward calculations show that, just by introducing static magnetic fields and gradients, no phase difference between the two superposition components can be observed at the end of a complete oscillation, i.e. when the interferometer closes.
In order to perform Ramsey interferometry to witness generation of spatial superposition, we introduce the possibility of tilting the $x$-axis of an angle $\theta_g$ in the $x-z$ plane, making the two ND superposition components sense a slightly different Earth gravitational potential.
In this scenario, the new reference frame coordinates will be (starting from $z=0$):
\begin{equation}
    \label{equ:NewFrame}
         x' = x\cos\theta_g \;\;\;\;\;\;\; z' = x\sin\theta_g\;.
\end{equation}
For an arbitrary $\theta_g\in(0,\pi/2)$, the two branches oscillates at different heights, thus accumulating a gravitational potential energy $mg\sin\theta$.
This can be encoded in the tilted Hamiltonian as
\begin{equation}
    \label{equ:TiltedHam}
    \mathcal{H} = \hbar\omega \had \ha +\hbar\Lambda [\hS]\left(\ha + \had\right) +\hbar\gamma_e B_0\hS +\hbar D\hS^2
\end{equation}
where $\Lambda[\hS] = \lambda_0 + \lambda_g + \lambda \hS$, and the coupling frequency $\lambda_g = (m g x_\text{zpf}/\hbar)\sin\theta_g$ (being $g$ the gravitational acceleration) accounts for the gravitational force contribution.
Having now introduced a coupling that involves only the CoM and not the spin, we now proceed to show how this induce a measurable difference of phase at the end of a period.
The mathematical treatment of the Hamiltonian in Eq. \eqref{equ:TiltedHam} is essentially equivalent to the one of the previous section.
First we expand it on the $\hS$ eigenbasis, allowing for the substitution of the spin operator with its eigenvalues $\hS\rightarrow s_j$:
\begin{equation}
    \mathcal{H} = \hbar\omega \had \ha +\hbar\Lambda_j\left(\ha + \had\right) +\hbar\gamma_e B_0s_j +\hbar Ds_j^2\;.
\end{equation}
The system eigenvalues and eigenvectors now read:
\begin{align}
        \ket{\psi'_j} &= D\left(-\frac{\Lambda_j}{\omega}\right)\ket{n}\;, \\
        E'_j &= \hbar\omega\left( n- \frac{\Lambda_j^2}{\omega^2} +\frac{\gamma_e B_0 s_j}{\omega}+\frac{D s_j^2}{\omega}\right)\;.
\end{align}
From this, we observe how $\lambda_g$ has the same effect as $\lambda_0$ to shift concurrently the center of oscillation of the two branches.
To explicitly express the difference of phase, we will make use of the updated time propagator
\begin{equation}
        \label{equ:TiltedTimePropagator}
        \hat{\mathcal{U}}_j(t) = D^\dagger\left(\frac{\Lambda_j}{\omega}\right) e^{-i\Phi_j t}e^{-i\omega \had\ha t}D\left(\frac{\Lambda_j}{\omega}\right)\;,
\end{equation}
with $\Phi_j = E'_j/\hbar$.

Assuming again the initial state in Eq. \eqref{psiI}, its generic time evolution under this new Hamiltonian can be expressed as:
\begin{align}
    \ket{\Psi(t)} =& \frac{1}{\sqrt{2}}\Bigg( e^{-i \vartheta_+(t)}\ket{\frac{\Lambda_+}{\omega}\left(e^{-i\omega t}-1\right)} \nonumber\\
    &+ e^{-i\vartheta_-(t)}\ket{ \frac{\Lambda_-}{\omega}\left(e^{-i\omega t}-1\right)}  \Bigg)\;,
\end{align}
where now
\begin{equation}
        \vartheta_j(t) = \Phi_j t +\left| \frac{\Lambda_j}{\omega} \right|^2\text{ sin}(\omega t) \;.
\end{equation}
After a full period of oscillation $t_1=2\pi/\omega$, a Ramsey scheme can be used to probe the phase difference
\begin{align}
\Delta\vartheta(t_1) =& \vartheta_+(t_1)-\vartheta_-(t_1) \nonumber\\
=& \frac{2\pi}{\omega}\left[\left(D+\gamma_eB_0-\frac{\Lambda_+^2}{\omega} \right) - \left(D-\gamma_eB_0-\frac{\Lambda_-^2}{\omega}\right)\right] \nonumber\\
 =& \frac{8\pi\lambda_g\lambda}{\omega^2}\;,
    %\begin{aligned}
%        \Delta\Theta(T) &= \Theta_+(T)-\Theta_-(T) = \frac{T}{\hbar}(E_+-E-)\\
%        & =T\left[   \left(-\frac{\Lambda_+^2}{\Omega}+\gamma_eB_0+D \right)  -\left( -\frac{\Lambda_-^2}{\Omega}+\gamma_eB_0+D\right)\right]\\
%        &=T\left[-\frac{1}{\Omega}\left(\Lambda_+ + \Lambda_- \right)\left(\Lambda_+ - \Lambda_-  \right) +2 \gamma_eB_0\right]\\
%        &=T\left[-\frac{1}{\Omega}(2\lambda_0+2\lambda_g)(2\lambda) + 2\gamma_eB_0\right]\\
%        &= T\left[-\cancel{\frac{4\lambda_0\lambda}{\Omega}} +\cancel{2\gamma_e\Omega} - \frac{4\lambda_g\lambda}{\Omega}\right]
%    \end{aligned}
\end{align}
which, expressed in terms of the system, variables reads
\begin{equation}
    \Delta\vartheta = -\frac{8\pi}{\omega^2}\lambda_g\lambda = -4\pi\left(\frac{\mu_0m}{\chi V} \right)^{3/2}\frac{\gamma_eg}{B'^2}\sin\theta_g\;.
\end{equation}
The only two free parameters on which such a phase difference depends are:
\begin{itemize}
    \item the tilting angle $\theta_g$: for $\theta_g=0$, no phase difference can be observed, since there is no difference of potential energy due to the Earth gravitational field;
    \item the amplitude of the magnetic field gradient $B'$ along the $x$ direction: a higher $B'$ corresponds to faster and smaller oscillations, with minor displacements compared to slower cases due to smaller $B'$.
\end{itemize}

\subsection{ND dynamics in presence of Dynamical Decoupling}

\begin{figure}[t]
	\begin{subfigure}[c]{0.48\textwidth}
		\centering
		\includegraphics[width=\textwidth]{phase_DD_NN4_B10-3.3new1.png}
		\subcaption{$B_0=5\times10^{-4}$ T; $\omega_{DD}=4\omega$.}
		\label{NN4bis}
	\end{subfigure}
	\hfill
	\begin{subfigure}[c]{0.48\textwidth}
		\centering
		\includegraphics[width=\textwidth]{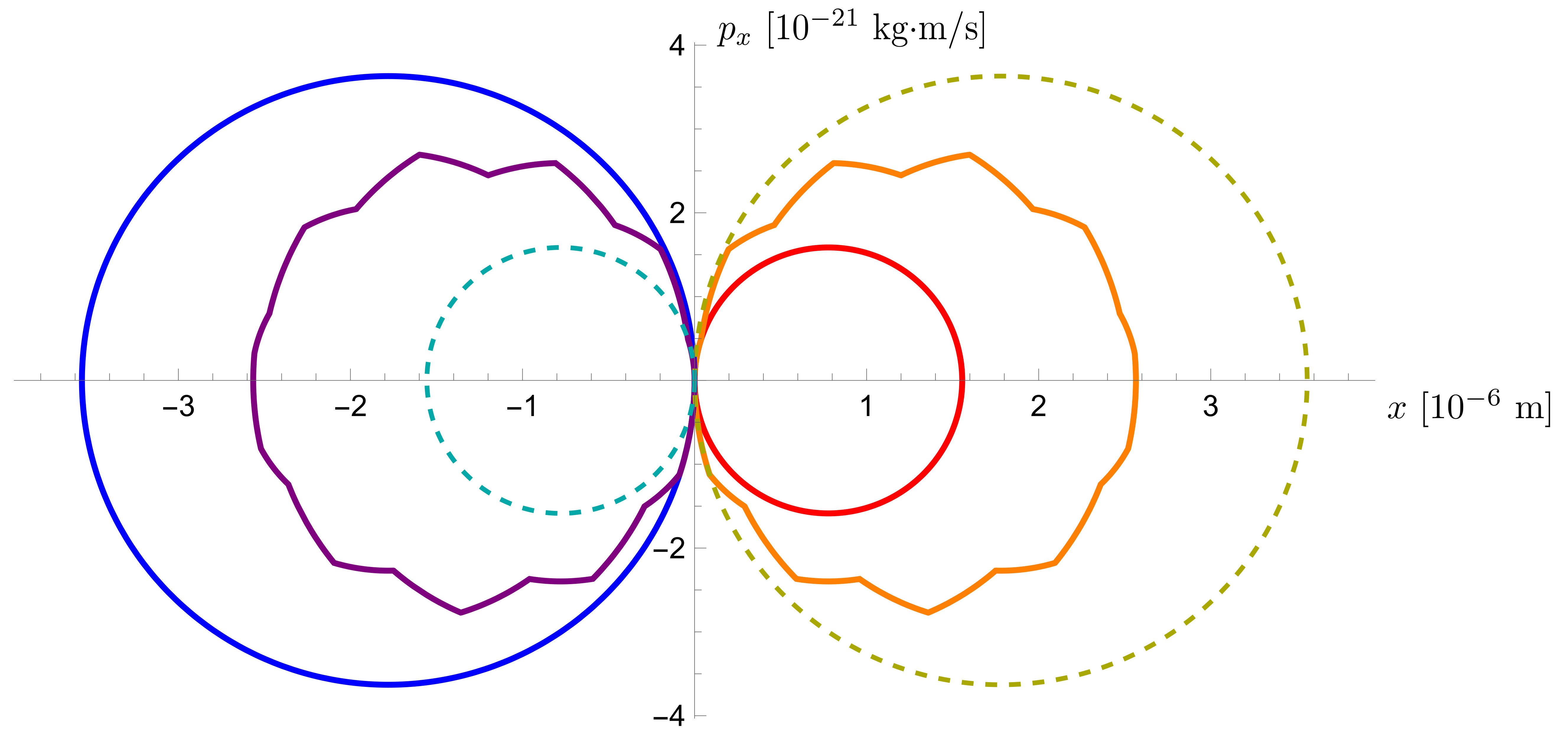}
		\subcaption{$B_0=5\times10^{-4}$ T; $\omega_{DD}=20\omega$.}
		\label{NN20}
	\end{subfigure}
	\hfill
	\begin{subfigure}[c]{0.48\textwidth}
		\centering
		\includegraphics[width=\textwidth]{phase_DD_NN200_B10-3.3new1.png}
		\subcaption{$B_0=5\times10^{-4}$ T; $\omega_{DD}=200\omega$.}
		\label{NN200bis}
	\end{subfigure}
	\caption{Plots (a-c): Phase space diagram of the ND superposition components, for different $\omega_{DD}$ values.
Red, green-dashed and orange curves, respectively: dynamics generated  by $H_x^{(\uparrow)}$, $H_x^{(\downarrow)}$ and $H_x^{(DD)}$ for $S_x=+1$.
%Red curve: dynamics generated by $H_x^{(\uparrow)}$ for $S_x=1$. Green dashed curve: dynamics generated by $H_x^{(\downarrow)}$ for $S_x=1$. Orange curve: dynamics generated by $H_x^{(DD)}$ for $S_x=1$.
Blue, azure-dashed and purple curves, respectively: dynamics generated by $H_x^{(\uparrow)}$, $H_x^{(\downarrow)}$ and $H_x^{(DD)}$ for $S_x=-1$.}
\label{phase_DD2}
\end{figure}

If one introduces DD \cite{Ped2020} by means of a train of microwave $\pi$ pulses flipping the $S_x$ sign at a frequency $\omega_{DD}=N \omega$, two scenarios are possible.
In the first one, the Hamiltonian in Eq. \eqref{equ:QHOHamiltonian} is kept time invariant by flipping the sign (i.e., the direction) of both $B_0$ and $B'$ (i.e. change both their signs) together with the $S_x$ sign flip, and the system dynamics remains the same.\\
Conversely, in the second one only the sign of $B'$ flips together with $S_x$, while $B_0$ remains the unperturbed.
Given the form of Eq. \eqref{ham}, this is equivalent to leaving $B'$ and $S_x$ untouched while flipping the sign of $B_0$ with frequency $\omega_{DD}$, leading to the time-dependent Hamiltonian in Eq. \eqref{hamDD}
%
%\begin{align}
%H_x^{(DD)}=&\sum_{i=0}^\infty\Bigg(\frac{p^2_x}{2m} + \frac{\chi V(B_0(-1)^i + B'x)^2}{2\mu_0} + \nonumber \\
% & +\hbar  \gamma_eS_x(B_0(-1)^i + B'x) + \hbar D S^2_x\Bigg)\cdot \nonumber\\
%&\cdot \left[\Theta\left(t-\frac{2\pi}{\omega_{DD}}i\right)-\Theta\left(t-\frac{2\pi}{\omega_{DD}}(i+1)\right)\right]  \;,
%\label{hamDD1}
%\end{align}
%
%\begin{equation}
%H_x^{(DD)}(t)=\sum_{j=0}^\infty\Bigg(\frac{p^2_x}{2m} + \frac{\chi V(B_0(-1)^j + B'x)^2}{2\mu_0} + \hbar  \gamma_eS_x(B_0(-1)^j + B'x) + \hbar D S^2_x\Bigg)\cdot \left[\Theta\left(t-\frac{2\pi j}{\omega_{DD}}\right)-\Theta\left(t-\frac{2\pi(j+1)}{\omega_{DD}}\right)\right]  \;,
%\label{hamDD1}
%\end{equation}
%
that, by defining $H_x^{(\uparrow)}\equiv H_x(B_0)$ and $H_x^{(\downarrow)}\equiv H_x(-B_0)$, can be written as:
%
%\begin{align}
%H_x^{(DD)}&=\sum_{k=0}^\infty\left\{H_x^{(\uparrow)}\left[\Theta\left(t-\frac{4\pi k}{\omega_{DD}}\right)-\Theta\left(t-\frac{2\pi(2k+1)}{\omega_{DD}}\right)\right]+\right.\nonumber\\
%&\left.+ H_x^{(\downarrow)}\left[\Theta\left(t-\frac{2\pi(2k+1)}{\omega_{DD}}\right)-\Theta\left(t-\frac{4\pi(k+1)}{\omega_{DD}}\right)\right]   \right\}  \;,
%\label{hamDD2}
%\end{align}
\begin{widetext}
\begin{equation}
H_x^{(DD)}(t)=\sum_{k=0}^\infty\left\{H_x^{(\uparrow)}\left[\Theta\left(t-\frac{4\pi k}{\omega_{DD}}\right)-\Theta\left(t-\frac{2\pi(2k+1)}{\omega_{DD}}\right)\right]+ H_x^{(\downarrow)}\left[\Theta\left(t-\frac{2\pi(2k+1)}{\omega_{DD}}\right)-\Theta\left(t-\frac{4\pi(k+1)}{\omega_{DD}}\right)\right] \right\}  \;,
\label{hamDD2}
\end{equation}
\end{widetext}
where $\Theta(x)$ indicates the Heaviside function.\\
Eq. \eqref{hamDD2} highlights how each ND superposition component will swap between the two Hamiltonians $H_x^{(\uparrow)}$ and $H_x^{(\downarrow)}$, both generating a (shifted) harmonic motion but with different equilibrium positions.
The global ND dynamics can be obtained as a composition of the dynamics in the single time intervals, resulting, for both superposition components, in a sort of forced oscillatory regime caused by the flipping of the harmonic oscillator equilibrium positions at each DD pulse.
This can be appreciated by applying the evolution operator $\hat{U}^{(DD)}=\exp(-\frac{i}{\hbar}\int_0^t\mathrm{d}t' H^{(DD)}_x(t))$ to the ND initial state; given the commutativity of the different terms in Eq. \eqref{hamDD2}, one has:
\begin{widetext}
\begin{equation}
%\ket{\Psi_I}\longrightarrow\hat{U}^{(DD)}\ket{\Psi_I}=\lim_{n\rightarrow\infty}\left(\hat{U}_{n}^{(\uparrow\downarrow)}\cdots \hat{U}_{2k+1}^{(\downarrow)}\hat{U}_{2k}^{(\uparrow)}\cdots \hat{U}_1^{(\downarrow)}\hat{U}_0^{(\uparrow)}\right)\ket{\Psi_I}\;,
\ket{\Psi_I}\longrightarrow\hat{U}^{(DD)}\ket{\Psi_I}=\lim_{n\rightarrow\infty}\left(\hat{U}_{2n+1}^{(\downarrow)}\hat{U}_{2n}^{(\uparrow)}\cdots \hat{U}_{2k+1}^{(\downarrow)}\hat{U}_{2k}^{(\uparrow)}\cdots \hat{U}_1^{(\downarrow)}\hat{U}_0^{(\uparrow)}\right)\ket{\Psi_I}\;,
%\frac{\ket{-1}\ket{\alpha(t)}_-^{(DD)}+\ket{+1}\ket{\alpha(t)}_+^{(DD)}}{\sqrt2} \;,
\label{evol1U}
\end{equation}
\end{widetext}
being:
\begin{align}
\hat{U}_{2k}^{(\uparrow)} ={} & \exp\left(-\frac{i}{\hbar}\int_0^t\mathrm{d}t' H^{(\uparrow)}_x \left[\Theta\left(t-\frac{4\pi k}{\omega_{DD}}\right) \right.\right. \nonumber \\
& \left.\left. -\Theta\left(t-\frac{2\pi(2k+1)}{\omega_{DD}}\right)\right]\right)\;; \label{evol1up}\\
\hat{U}_{2k+1}^{(\downarrow)} ={} & \exp\left(-\frac{i}{\hbar}\int_0^t\mathrm{d}t' H^{(\downarrow)}_x \left[\Theta\left(t-\frac{2\pi(2k+1)}{\omega_{DD}}\right) \right.\right. \nonumber \\
& \left.\left. -\Theta\left(t-\frac{4\pi(k+1)}{\omega_{DD}}\right)\right]\right)\;. \label{evol1down}
\end{align}
By applying such an evolution to the ground state $\ket*0$ of a harmonic oscillator one obtains a state of the same kind as the one in Eq. \eqref{evol}:
\begin{equation}
\ket{\Psi_I}\longrightarrow\hat{U}^{(DD)}\ket{\Psi_I}=\frac{\ket{-1}\ket{\alpha(t)}_-^{(DD)}+\ket{+1}\ket{\alpha(t)}_+^{(DD)}}{\sqrt2} \;,
\label{evol1DD}
\end{equation}
with
%
%\begin{align}
%\ket{\alpha(t)}_\pm^{(DD)}=&\sum_{k=0}^\infty\left\{\mathcal{C}_{2k}(t)\left[\Theta\left(t-\frac{4\pi k}{\omega_{DD}}\right)-\Theta\left(t-\frac{2\pi(2k+1)}{\omega_{DD}}\right)\right]\ket{\alpha_{2k}(t)}_\pm+\right. \nonumber\\ &\left.+\mathcal{C}_{2k+1}(t)\left[\Theta\left(t-\frac{2\pi(2k+1)}{\omega_{DD}}\right)-\Theta\left(t-\frac{4\pi(k+1)}{\omega_{DD}}\right)\right]\ket{\alpha_{2k+1}(t)}_\pm \right\}  \;.
%\label{evol2DD}
%\end{align}
\begin{align}
%\ket{\alpha(t)}_\pm^{(DD)} =& \sum_{j=0}^\infty\mathcal{C}^{(j)}_\pm(t)\left[\Theta\left(t-\frac{2\pi j}{\omega_{DD}}\right)\right.\nonumber\\
%\left. &-\Theta\left(t-\frac{2\pi(j+1)}{\omega_{DD}}\right)\right]\ket{\alpha^{(j)}_\pm(t)}  \;,
%\label{evol2DD}
\ket{\alpha(t)}_\pm^{(DD)} =& \sum_{j=0}^\infty \mathcal{C}^{(j)}_\pm(t) \left[ \Theta\left(t-\frac{2\pi j}{\omega_{DD}}\right) \right. \nonumber \\
& \left. - \Theta\left(t-\frac{2\pi(j+1)}{\omega_{DD}}\right) \right] \ket{\alpha^{(j)}_\pm(t)} \;, \label{evol2DD}
\end{align}
where $\mathcal{C}^{(j)}_\pm(t)$ plays the role of a recursive phase coefficient ($|\mathcal{C}^{(j)}_\pm(t)|^2=1$) and $\ket{\alpha^{(j)}_\pm(t)}$ is the coherent state with amplitude $|\alpha^{(j)}_\pm(t)|^2$ appearing in the time interval $t\in\left[\frac{2\pi j}{\omega_{DD}};\frac{2\pi(j+1)}{\omega_{DD}}\right]$.\\
By introducing the coefficients $\lambda_\pm^{(j)}=(-1)^j\lambda_0\pm\lambda$, with some algebra one obtains:
%
%\begin{equation}
%%\ket{\alpha^{(j)}_\pm(t)} = \ket{\frac{\lambda_\pm^{(j)}}{\omega}\left(1-e^{i\omega t}\right)+\sum_{\xi=1}^{j\geq1}\left(\frac{\lambda_\pm^{(\xi-1)}}{\omega}\left(1-e^{\frac{i2\pi\omega}{\omega_{DD}}\xi}\right)\right)}
%\ket{\alpha^{(j)}_\pm(t)} = \ket{\frac{\lambda_\pm^{(j)}\left(e^{i\omega t}-1\right)}{\omega}+\sum_{l=1}^{j\geq1}\left(\frac{\lambda_\pm^{(l-1)}}{\omega}\left(e^{\frac{i2\pi\omega}{\omega_{DD}}l}-1\right)\right)}
%\label{alphaDD_j}
%\end{equation}
%
\begin{align}
%\alpha_\pm^{(j)}(t) =& \frac{e^{-i\frac{2\pi\omega}{\omega_{DD}}}-1}{\omega} \left(\sum_{l=0}^{(j-2)\geq0} \lambda_\pm^{(l)} e^{-i\frac{\omega(j-l-2)}{\omega_{DD}}}\right) e^{-i\omega\left(t-\frac{2\pi(j-1)}{\omega_{DD}}\right)} \nonumber\\ &+ \frac{\lambda_\pm^{(j-1)}}{\omega} \left( e^{-i\omega\left(t-\frac{2\pi(j-1)}{\omega_{DD}}\right)} - 1 \right)
\alpha_\pm^{(j)}(t) =& e^{-i\omega\left(t-\frac{2\pi j}{\omega_{DD}}\right)} \left(e^{-i\frac{2\pi\omega}{\omega_{DD}}}-1\right) \sum_{l=0}^{(j-1)\geq0} \frac{\lambda_\pm^{(l)}}{\omega} e^{-i\frac{2\pi\omega(j-l-1)}{\omega_{DD}}} \nonumber\\
& + \frac{\lambda_\pm^{(j)}}{\omega} \left( e^{-i\omega\left(t-\frac{2\pi j}{\omega_{DD}}\right)} - 1 \right)
\label{alphaDD_j}
\end{align}
and
%
%\begin{align}
%\mathcal{C}^{(j)}_\pm(t)=& \mathcal{C}_\pm^{(j-1)}\left(t=\frac{2\pi(j-1)}{\omega_{DD}}\right)\cdot \exp\left\{-i\theta_\pm^{(j)}\left(t-\frac{2\pi j}{\omega_{DD}}\right)+\right. \nonumber\\
%& \left. + i\frac{\lambda_\pm^{(j)}}{\omega}\Im\left[\alpha_\pm^{(j)}\left(t-\frac{2\pi j}{\omega_{DD}}\right)\left(1-e^{-i\omega\left(t-\frac{2\pi j}{\omega_{DD}}\right)}\right)\right]\right\}\;,
%%
%%e^{i\omega\left[ t\left(\left|\frac{\lambda_\pm^{(j)}}{\omega}\right|^2-\frac12\right)+ \sum_{l=1}^{j\geq1}\frac{2\pi l}{\omega_{DD}}\left(\left|\frac{\lambda_\pm^{(l-1)}}{\omega}\right|^2-\frac12\right)\right]}\cdot\nonumber\\ &\cdot\exp\left[i\left(\varphi_\pm^{(j)}(t)+\sum_{l=1}^{j\geq1}\varphi_\pm^{(l)}\left(t=\frac{2\pi}{\omega_{DD}}l\right)\right)\right] \;,
%\label{C_j}
%\end{align}
%%
%with $\mathcal C_\pm^{(0)}=1$ and
%\begin{equation}
%\theta_\pm^{(j)}(t)= \left(D\pm\gamma_eB_0-\frac{(\lambda_\pm^{(j)})^2}{\omega}\right)t+\left(\frac{\lambda_\pm^{(j)}}{\omega}\right)^2\sin(\omega t) \;.
%\end{equation}
%
%and
%
\begin{equation}
\mathcal{C}^{(j)}_\pm(t)= \exp\left[-i\sum_{l=0}^{(j-1)\geq0}\mathcal Q_\pm^{(l)}\left(\frac{2\pi}{\omega_{DD}}\right)-i\mathcal Q_\pm^{(j)}\left(t-\frac{2\pi j}{\omega_{DD}}\right)\right]\;,
%
%e^{i\omega\left[ t\left(\left|\frac{\lambda_\pm^{(j)}}{\omega}\right|^2-\frac12\right)+ \sum_{l=1}^{j\geq1}\frac{2\pi l}{\omega_{DD}}\left(\left|\frac{\lambda_\pm^{(l-1)}}{\omega}\right|^2-\frac12\right)\right]}\cdot\nonumber\\ &\cdot\exp\left[i\left(\varphi_\pm^{(j)}(t)+\sum_{l=1}^{j\geq1}\varphi_\pm^{(l)}\left(t=\frac{2\pi}{\omega_{DD}}l\right)\right)\right] \;,
\label{C_j}
\end{equation}
with
\begin{align}
\mathcal Q_\pm^{(k)}(t)=& \left(D\pm\gamma_eB_0-\frac{(\lambda_\pm^{(k)})^2}{\omega}\right)t+\left(\frac{\lambda_\pm^{(k)}}{\omega}\right)^2\sin(\omega t) \nonumber\\
&+ \frac{\lambda_\pm^{(k)}}{\omega}\Im\left[\alpha_\pm^{(k)}(t)\left(1-e^{-i\omega t}\right)\right]
 \;.
\end{align}
%
%\begin{align}
%\mathcal{C}^{(j)}_\pm(t)=& e^{i\omega\left[ t\left(\left|\frac{\lambda_\pm^{(j)}}{\omega}\right|^2-\frac12\right)+ \sum_{l=1}^{j\geq1}\frac{2\pi l}{\omega_{DD}}\left(\left|\frac{\lambda_\pm^{(l-1)}}{\omega}\right|^2-\frac12\right)\right]}\cdot\nonumber\\ &\cdot\exp\left[i\left(\varphi_\pm^{(j)}(t)+\sum_{l=1}^{j\geq1}\varphi_\pm^{(l)}\left(t=\frac{2\pi}{\omega_{DD}}l\right)\right)\right] \;,
%\label{C_j}
%\end{align}
%%
%being
%%
%\begin{align}
%\varphi^{(j)}_\pm(t)=& -i\left[\left|\frac{\lambda_\pm^{(j)}}{\omega}\right|^2\left(1-e^{i\omega t}\right) \right. \nonumber \\
%&+ \sum_{l=1}^{j\geq1}\left(\left|\frac{\lambda_\pm^{(l-1)}}{\omega}\right|^2\left(1-e^{\frac{i2\pi\omega}{\omega_{DD}}l}\right)\right) \nonumber \\
%&+\frac12\left|\frac{\lambda_\pm^{(j)}}{\omega}\left(1-e^{i\omega t}\right) +\sum_{l=1}^{j\geq1}\frac{\lambda_\pm^{(l-1)}}{\omega}\left(1-e^{\frac{i2\pi\omega}{\omega_{DD}}l}\right)\right|^2 \nonumber \\
%&- \left. \frac12\left|\sum_{l=1}^{j\geq1}\frac{\lambda_\pm^{(l-1)}}{\omega}\left(1-e^{\frac{i2\pi\omega}{\omega_{DD}}l}\right)\right|^2 \right] \;.
%\label{phi_j}
%\end{align}
%
The dynamics of the ND superposition components in presence of DD is highlighted by the phase space diagram in Fig. \ref{phase_DD2}, obtained considering $B_0=5\times10^{-4}$ T and different $N$ values for $\omega_{DD}=N\omega$.
The figure shows how, for small $N$, the ``jumps'' between $H_x^{(\uparrow)}$ and $H_x^{(\downarrow)}$ appear in both trajectories (orange and purple curves for $S_x=+1$ and $S_x=-1$, respectively), approaching the one of a harmonic oscillator with an ``average'' equilibrium position for large $N$ (i.e., in the actual DD scenario).
Remarkably, even though a non-negligible $B_0$ is present, the trajectories dictated by $H_x^{(DD)}$ are symmetric with respect to the origin, as it was before for $B_0=0$ (Fig. \ref{strong}).
This means that DD makes our ND superpositions unperturbed by residual (unwanted) bias fields in our setup, avoiding the need of counterbalancing them or include them in the theoretical model.
%Remarkably, even though the non-negligible $B_0$ considered here already produces a phase difference between the two ND superposition components without DD (or for $B_0$ flipping accordingly with $\omega_{DD}$), the trajectories dictated by $H_x^{(DD)}$ are symmetric with respect to the origin, as it was before for $B_0=0$ (Fig. \ref{strong}).
%This means that, with DD, one needs to flip also $B_0$ to observe some $B_0$-induced phase difference between the ND superposition components; at the same time, this means that DD makes our ND superpositions unperturbed by residual (unwanted) bias fields in our setup.

%\appendix
\section{Magnetic field configuration and sensitivity analysis for the ND interferometer}
\begin{figure*}[t]
	\centering
    \includegraphics[width=0.9\textwidth]{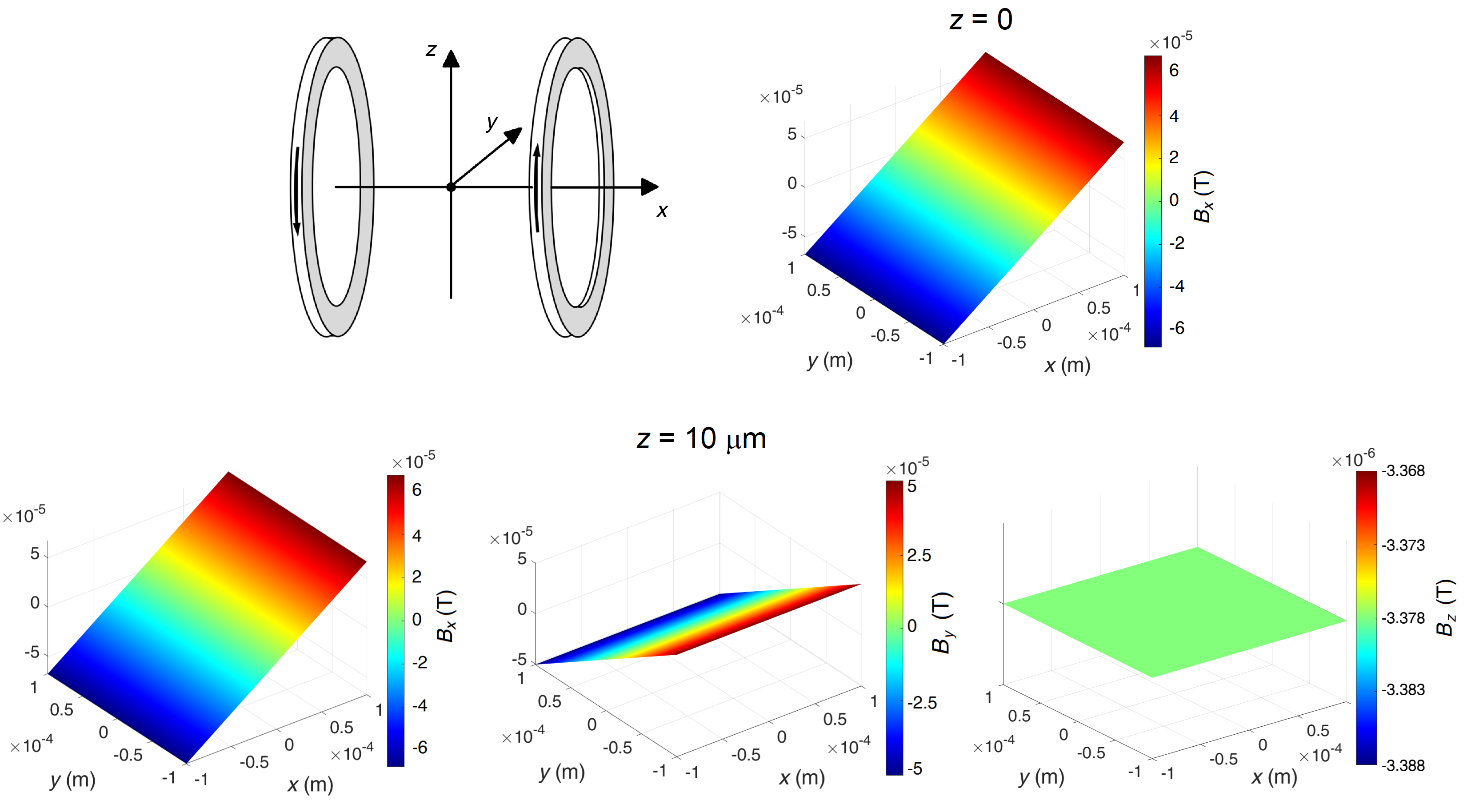}
%	\begin{subfigure}[c]{\textwidth}
%		\centering
%		\includegraphics[width=0.7\textwidth]{Immagine1.png}
%	\end{subfigure}
%	
%	\vspace{0.5cm} % spazio verticale tra le immagini
%	
%	\begin{subfigure}[c]{\textwidth}
%		\centering
%		\includegraphics[width=0.8\textwidth]{Immagine2a.png}
%	\end{subfigure}
	
	\caption{Maps of the magnetic field components in the $x$-$y$ planes for $z=0$ and $z=10\,\mathrm{\mu m}$, generated by the anti-Helmoltz coil configuration sketched in the top-left panel, for a magnetic field gradient (along $x$) $B'=0.663$ T/m.
The top-right panel shows the linear spatial distribution of the magnetic field component along $x$ ($B_x$) at $z=0$, where both the field components along $y$ and $z$ ($B_y$ and $B_z$, respectively) are null.
The three maps in the bottom panels show the $B_x$, $B_y$ and $B_z$ components, evaluated at $z=10$ $\mu$m; the linear variations of $B_x$ and $B_y$ with $x$ and $y$, respectively, can be observed, while $B_z$ can be considered uniform in the selected range.}
	\label{fig:mag}
\end{figure*}

\begin{figure*}[t]
	\centering
    \includegraphics[width=\textwidth]{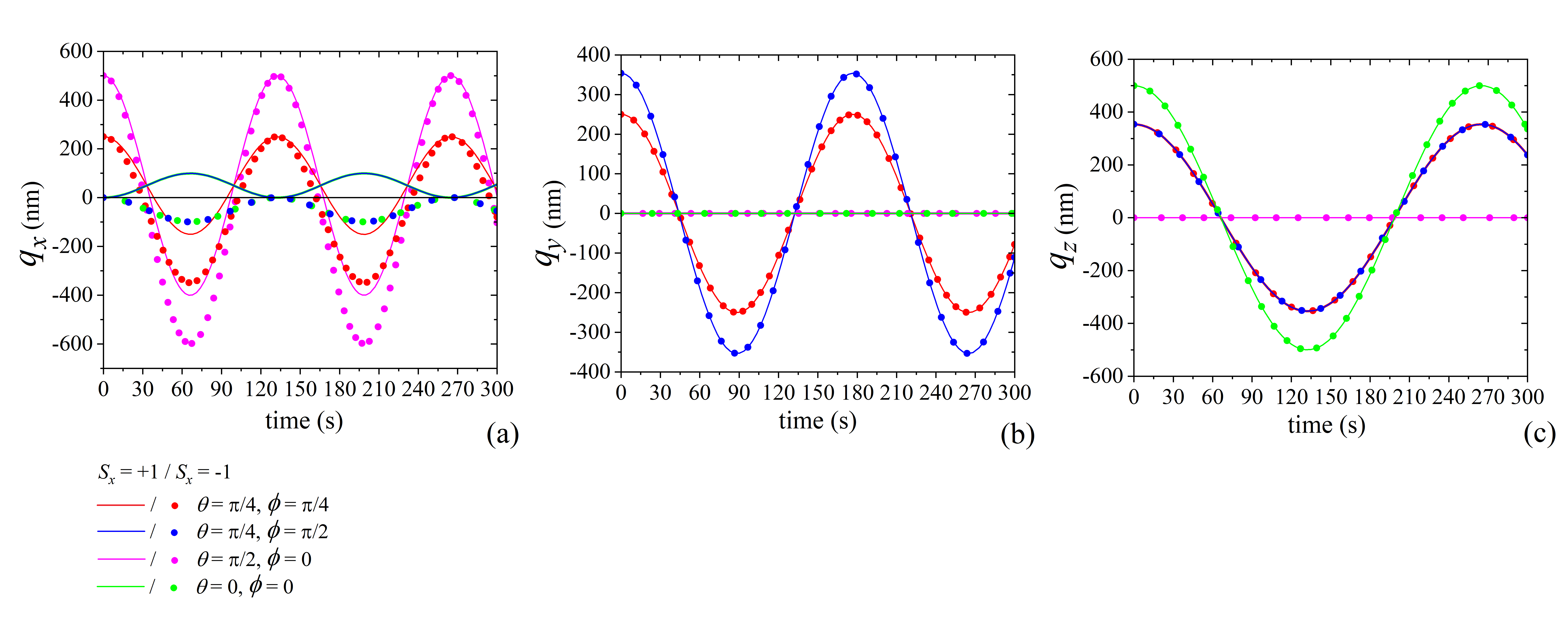}
	%\begin{subfigure}[t]{0.5\textwidth}
%		\centering
%		\includegraphics[width=\textwidth]{figure5a.png}
%	\end{subfigure}
%	
%	%\hspace{0.5cm} % spazio orizzontale tra le immagini
%	
%	\begin{subfigure}[t]{0.5\textwidth}
%		\centering
%		\includegraphics[width=\textwidth]{figure5b.png}
%	\end{subfigure}
%
%    %\hspace{0.5cm} % spazio orizzontale tra le immagini
%	
%	\begin{subfigure}[t]{0.5\textwidth}
%		\centering
%		\includegraphics[width=\textwidth]{figure5c.png}
%	\end{subfigure}

	\caption{Analysis of the ND superposition components trajectories in the 3D space for different initial positions (expressed in spherical coordinates with radial distance fixed at 500 nm). Solid (dotted) curves indicate the trajectories of the $S_x=+1$ ($S_x=-1$) ND component.  Plot (a): trajectories along $x$. Plot (b): trajectories along $y$. Plot (c): trajectories along $z$. %In the $x$ and $y$ directions, the orange curves overlap with the azure ones; in the $z$ direction, the azure and magenta curves coincide.
}
	\label{fig:XX}
\end{figure*}

\begin{figure}[t]
	\centering
		\centering
		\includegraphics[width=0.9\columnwidth]{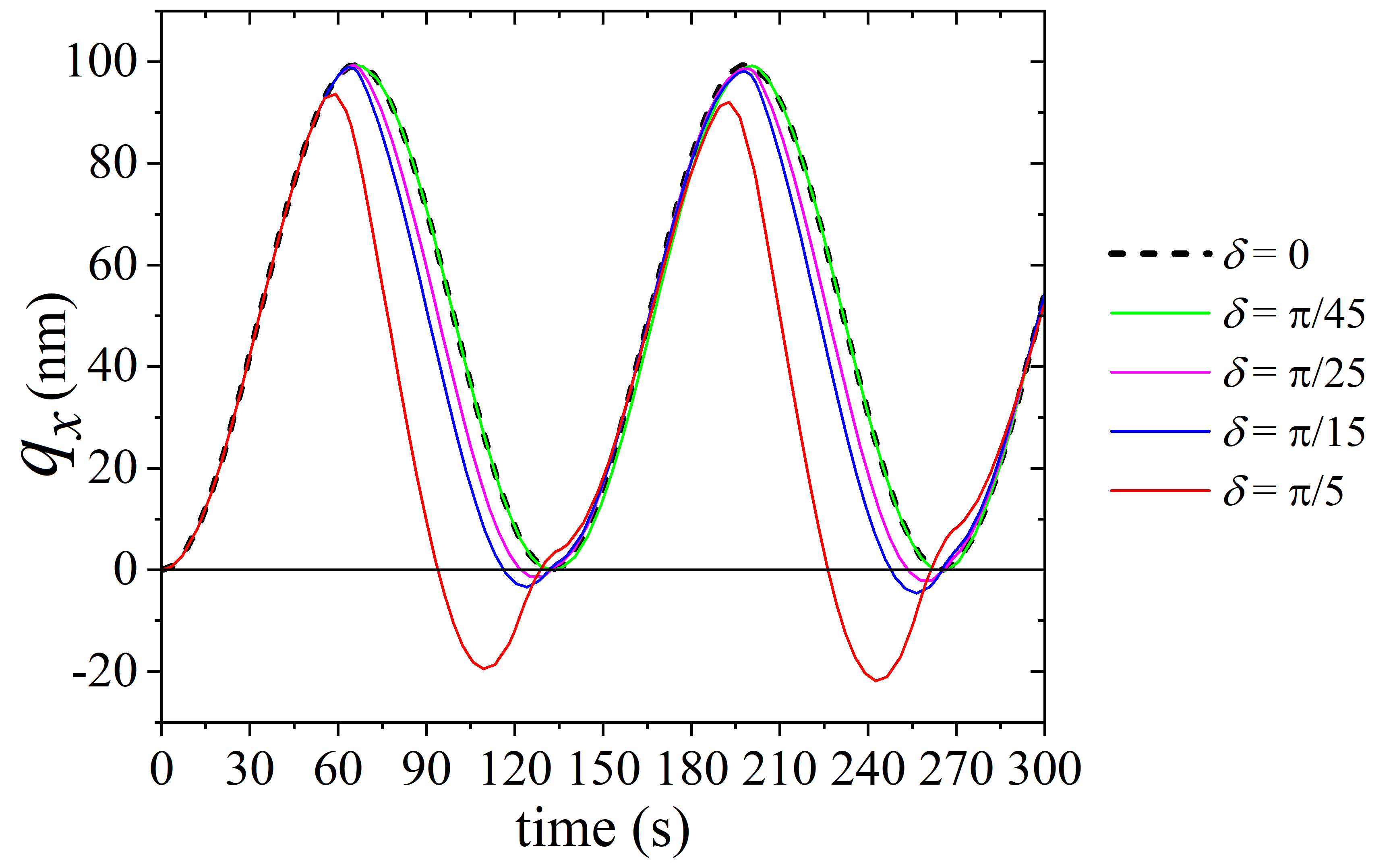}
	\caption{Comparison of one full oscillation of the $S_x=+1$ ND superposition component along the $x$-axis, with the axes origin as the ND initial position and both $S_x$ and $B'$ flipping sign with frequency $\omega_{DD}= N\omega$ (with $N=200$), but with an eventual phase shift $\delta=0,\pi/5,\pi/15,\pi/25,\pi/45$ between them.}
	\label{fig:YY}
\end{figure}

The considered magnetic field spatial distribution can be generated with an anti-Helmholtz coil, a field source configuration typically used for magnetic trapping of atoms \cite{migdall1985}.
With this configuration, which consists of two coaxial coils currying equal currents in the opposite directions, it is possible to produce a field that is zero at the center and increases linearly away from the center.
As a proof, we have also performed numerical simulations of the ND trajectory, where the magnetic field produced by the anti-Helmholtz coil has been evaluated by solving the Biot-Savart law.
The assumption of nearly static field is considered, being a good approximation for RF coils with size much lower than the vacuum wavelength.\\
For a single circular coil approximated as a loop with cross-section parallel to the $y-z$ plane, having radius $r_c$ and centered at $x = x_c$, the components of the magnetic field are expressed as
\begin{equation}
	\begin{split}
		B_x&=\frac{\mu_0\mathfrak{F}}{2\pi\alpha^2\beta}\left[\left({r_c}^2-r^2\right)\mathrm{E}\left(k^2\right)+\alpha^2\mathrm{K}\left(k^2\right)\right],\\ B_y&=\frac{\mu_0\mathfrak{F}(x-x_c)y}{2\pi\alpha^2\beta\rho^2}\left[\left({r_c}^2+r^2\right)\mathrm{E}\left(k^2\right)-\alpha^2\mathrm{K}\left(k^2\right)\right],\\
		B_z&=\frac{z}{y}B_y, \\
	\end{split}\label{app:mag1}
\end{equation}
where $\mu_0$ is the magnetic permeability of free space, $\mathfrak{F}$ is the magnetomotive force, $\mathrm{K}$ and $\mathrm{E}$ are the elliptic integrals of the first and second kind, $r^2=(x-x_c)^2+y^2+z^2$, $\rho^2=y^2+z^2$, $\alpha^2={r_c}^2+r^2-2r_c\rho$, $\beta^2={r_c}^2+r^2+2r_c\rho$ and $k^2=1-\frac{\alpha^2}{\beta^2}$ \cite{chu1998,garrett1963}.
The field generated by the anti-Helmholtz coil is obtained by adding the contributions of two coaxial coils separated by distance $d_c$.\\
As an example, to produce a magnetic field with a uniform gradient ($B'$) of $0.663\,\mathrm{T/m}$, we can use two coils with $r_c=3\,\mathrm{cm}$ and $d_c=3\,\mathrm{cm}$, supplied with $564\,\mathrm{At}$.
Fig. \ref{fig:mag} shows the maps of the magnetic field components in the $x$-$y$ planes for $z=0$ and $z=10\,\mathrm{\mu m}$.\\
To support our results, we perform a sensitivity analysis by means of a numerical model that simulates the ND trajectories in the $x,y,z$ directions, under the effect of the magnetic force due to the external magnetic field, generated with the anti-Helmholtz coil.
In this analysis, we consider the components of the magnetic field and its gradient along all the three axial directions, thus including 3D force contributions.
For simplicity, we assume that the magnetic trap is able to compensate the effect of the gravitational force, which is not included in the magneto-mechanical model. Specifically, we solve the following system of equations:
\begin{equation}
		m\frac{d}{dt}\bm{v}=\left(\bm{\mu}\cdot\nabla\right)\bm{B} \;\;\;\;\;\; \frac{d}{dt}\bm{q}=\bm{v},
	\label{app:mag2}
\end{equation}
where $m$ is the mass of the ND, $\bm{v}$ its velocity, $\bm{q}$ its position, $\bm{\mu}$ its magnetic momentum, defined as the superposition of the diamagnetic contribution $\frac{\chi V\bm{B}}{\mu_0}$ (with $\chi$ and $V$ being the ND susceptibility and volume, respectively) and the spin contribution $\mu_B S_x \hat{\bm{x}}$, assumed to be different from zero only along the $x$-axis.
The magnetic field $\bm{B}$ is calculated according to Eq. \eqref{app:mag1}, and the DD is implemented by flipping the $S_x$ spin component synchronously with the applied magnetic field gradient $B'$.\\
The ND trajectory is numerically simulated by solving Eq.s \eqref{app:mag2} by means of the explicit and adaptive-step-size Runge-Kutta 45 method.
%, based on the Dormand-Prince pair \cite{dormand1980}.\\
In the sensitivity analysis, we vary the coil magnetomotive force (and thus the magnetic field and its gradient) and the ND initial position.
Moreover, we analyze what happens when the MW $\pi$ pulse flipping the $S_x$ sign of the ND superposition components is not exactly synchronous with the square waveform flipping $B'$.
%Figures \ref{fig:XX}(a)-(c) and \ref{fig:YY} present the results obtained for a magnetic field with a uniform gradient $B'=0.663$ T/m along $x$.
%This configuration is obtained by considering two coaxial coils with $r_c=3$ cm and spaced $3$ cm apart, each supplied with a current corresponding to $\mathfrak{F}=564$ At.
%The motion of the ND, assumed here to have a mass $m=5.6\times10^{-14}$ kg and with $S_x$ flipped synchronously with $B'$ at the DD frequency  $\omega_{DD}= N\omega$ (with $N=200$), is simulated by numerically integrating the system of equations \eqref{app:mag2} using the explicit forward Euler method.\\
Figures \ref{fig:XX}(a)-(c) and \ref{fig:YY} present the results obtained for a magnetic field with a uniform gradient $B'=0.663$ T/m along $x$, assuming that the ND mass is $m=5.6\times10^{-14}$ kg and $S_x$ is flipped synchronously with $B'$ at the DD frequency  $\omega_{DD}= N\omega$ (with $N=200$).\\
Specifically, Fig. \ref{fig:XX} shows how the trajectory of the ND varies for different initial positions, defined within the first octant of the 3D space in spherical coordinates.
The radial distance $r$ is fixed at 500 nm, while the angular coordinates $\theta$ and $\phi$ (being $x=r_c\sin\theta\cos\phi$, $y=r_c\sin\theta\sin\phi$ and $z=r_c\cos\theta$) vary in the $[0,\pi/2]$ interval.
Solid lines refer to the ND superposition component with $S_x = 1$, while the dashed lines correspond to the one with $S_x=-1$.
We can notice how the $S_x$ spin component only affects the trajectory along $x$, as proved by the overlap of the curves corresponding to $S_x=\pm1$ in Fig.s \ref{fig:XX}(b) and \ref{fig:XX}(c).
The $y$- and $z$-components of the ND trajectory exhibit independent behavior, characterized by oscillatory motion when the initial position coordinate is $y,z\neq0$.
Conversely, if the initial value of either coordinate is zero, no motion along that direction occurs.
Anyway, the absence of dependence on $S_x$ of the ND trajectories along the $y$ and $z$ directions allows us to ``decouple'' those spatial motions from the one of interest, focusing just on the $x$ direction.\\
Figure \ref{fig:YY}, instead, illustrates how the trajectory of the ND superposition component with $S_x=1$ and initial position $x=y=z=0$ changes when the DD implementation is not ideal, i.e., when the spin flip is not perfectly synchronized with $B'$.
As an example, we analyze the effect of a phase shift $\delta=\pi/n$ in flipping ($n =5,15,25,45$).
Compared to the perfectly-synchronized case (dashed black line), small $\delta$ values like, e.g. $\delta=\pi/25$, do not practically impact on the ND trajectory.
Conversely, as $\delta$ increases, the ND trajectory progressively moves away from the original path, and the harmonic oscillation is gradually lost.
Anyway, it is worth noticing that, up to $\delta=\pi/15$ (way above the confidence we can expect from the synchronization between the $S_x$ and $B'$ flips), such an effect remains still negligible, leaving the system dynamics basically unchanged.

\bibliographystyle{apsrev4-2}
\bibliography{ref}

\end{document}